\documentclass[aps,prb,twocolumn,superscriptaddress,longbibliography]{revtex4-2}

\usepackage{graphicx}
\usepackage[normalem]{ulem}
\usepackage{color}
\usepackage{sidecap}
\usepackage{xspace}

\usepackage{amsmath}
\usepackage{siunitx}
\usepackage{miller}

\usepackage[version=3]{mhchem}
\usepackage{natbib}
\usepackage[latin1]{inputenc}

\newcommand{\sroo}{Sr$_{2}$RuO$_{4}$}

\newcommand{\sro}{SrRuO$_{3}$}

\newcommand{\tc}{$\text{T}_\text{C}$}

\newcommand{\soc}{spin-orbit coupling}

\newcommand{\tas}{triple-axis spectrometer}
\newcommand{\tof}{time-of-flight}

\newcommand{\etal}{\textit{et al.}}

\begin{document}


\title{Magnon dispersion in ferromagnetic SrRuO$_3$}

\author{K. Jenni}
\affiliation{$I\hspace{-.1em}I$. Physikalisches Institut,
Universit\"at zu K\"oln, Z\"ulpicher Str. 77, D-50937 K\"oln,
Germany}

\author{S. Kunkem\"oller}
\affiliation{$I\hspace{-.1em}I$. Physikalisches Institut,
Universit\"at zu K\"oln, Z\"ulpicher Str. 77, D-50937 K\"oln,
Germany}

\author{A. Tewari}
\affiliation{$I\hspace{-.1em}I$. Physikalisches Institut,
Universit\"at zu K\"oln, Z\"ulpicher Str. 77, D-50937 K\"oln,
Germany}

\author{R. A. Ewings}
\affiliation{ISIS Pulsed Neutron and Muon Source, STFC Rutherford Appleton Laboratory, Harwell Campus, Didcot, Oxon, OX11 0QX, United Kingdom}

\author{Y. Sidis}
\affiliation{Universit\'e Paris-Saclay, CNRS, CEA, Laboratoire L\'eon Brillouin, 91191, Gif-sur-Yvette, France.}

\author{A. Schneidewind}
\affiliation{J\"ulich Centre for Neutron Science (JCNS) at Heinz Maier-Leibnitz Zentrum (MLZ), Forschungszentrum J\"ulich GmbH, Lichtenbergstraße 1, 85748 Garching, Germany}

\author{P. Steffens}
\affiliation{Institut Laue Langevin,71 avenue des Martyrs, 38000 Grenoble, France}

\author{A. A. Nugroho}
\affiliation{ Faculty of Mathematics and Natural Science, Institut Teknologi Bandung, Jalan Ganesha 10, 40132 Bandung, Indonesia}

\author{M. Braden}\email[e-mail: ]{braden@ph2.uni-koeln.de}
\affiliation{$I\hspace{-.1em}I$. Physikalisches Institut,
Universit\"at zu K\"oln, Z\"ulpicher Str. 77, D-50937 K\"oln,
Germany}





\date{\today}

\begin{abstract}

The magnetic excitations in ferromagnetic SrRuO$_3$ were studied by inelastic neutron scattering
combining experiments on triple-axis and time-of-flight spectrometers with and without polarization
analysis. A quadratic spin-wave dispersion with an anisotropy gap describes the low-energy
low-temperature response. The magnon dispersion extends to at least 35\,meV and there is no direct
evidence for a continuum of Stoner excitations below this energy. However, the magnon response is
weakened at higher energy. In addition to the anomalous softening of the spin-wave stiffness and of the gap, which
is induced by the topology of the Bloch states, the magnon excitations are broadened in energy and this effect increases upon heating.

\end{abstract}


\maketitle


\section{Introduction}

Among the Ruddlesden-Popper ruthenates Sr$_{n+1}$Ru$_{n}$O$_{3n+1}$,
SrRuO$_3$ is the only simple material to exhibit ferromagnetic order at zero magnetic field \cite{Randall1959,Jones1989,Chakoumakos1998}.
This ferromagnetism inspired the proposition of $p$-wave superconductivity in \sroo\  \cite{Rice1995,Baskar1996} with a pairing mechanism involving ferromagnetic fluctuations \cite{Steffens2019}.
But the magnetism in \sro\ itself is intriguing because of the connection to anomalies in various properties.
At the ferromagnetic transition temperature of \tc\ $=165\,\text{K}$ there is a kink in the direct-current transport measurement \cite{Allen1996}.
In addition the cell volume does not shrink in the ordered phase, which is known as the invar effect \cite{Kiyama1996}.
The spin degree of freedom thus seems to be coupled to charge and lattice degrees of freedom \cite{Allen1996,Klein1996,Kiyama1996}.
SrRuO$_3$ can be categorized as a 'bad metal', because the high-temperature resistivity passes through the Ioffe-Regel limit around 500\,K without indication of saturation \cite{Klein1996}.
In metallic magnets the question about the local or itinerant character is always challenging \cite{Moriya1985}.
Based on a pressure study of \tc , the SrRuO$_3$ material has been classified as a moderately weak itinerant ferromagnet \cite{Neumeier1994}, while
a recent ARPES study proposes a dual nature for majority and minority states in SrRuO$_3$ \cite{Hahn2021}.
According to itinerant Stoner theory, the low-$q$ spin-wave dispersion corresponds to a bound state and passes into a continuum of electron-hole pair excitations \cite{Blundell2001,Moriya1985} that so far has not been reported for SrRuO$_3$.

Magnetization measurements reveal a large anisotropy with the magnetic easy axis pointing along the elongation of the RuO$_6$ octahedron (orthorhombic $c$ axis in space group \textit{Pnma}) \cite{Kunkemoeller2017b}.
The anisotropy field of $\sim$10\,T documents strong \soc\ in this material \cite{Kanbayasi1976,Cao1997b,Kunkemoeller2016,Kunkemoeller2017b}.
This strong \soc\ also implies anomalous magnetotranport properties that can be attributed to Weyl points in the electronic structure \cite{Fang2003,Onoda2008,Itoh2016,Takiguchi2020,Kaneta-Takada2022}.
For SrRuO$_3$ the relation between the intrinsic anomalous Hall effect and the topology of the electronic structure was demonstrated for the
first time \cite{Fang2003}. The combination of orbital band degeneracy, magnetic exchange splitting and
spin-orbit coupling induces Weyl points and a Berry phase \cite{Itoh2016}, which are accepted to explain the peculiar temperature dependence of the
anomalous Hall effect \cite{Izumi1997,Fang2003,Koster2012,Itoh2016,Jenni2019}. Other anomalous magneto-transport properties corroborated the strong impact of Weyl points in SrRuO$_3$ \cite{Takiguchi2020,Kaneta-Takada2022,Kar2023}.

Previous inelastic neutron scattering (INS) studies on the magnetic excitations in SrRuO$_3$ focussed on the temperature dependencies of the magnon gap, $\Delta$, and of the spin-wave stiffness constant $D$ \cite{Itoh2016,Jenni2019}.
Single-crystal studies find anomalous temperature dependencies for both parameters that
were attributed to the impact of the Bloch states topology \cite{Jenni2019}. The Weyl points lead to an interconnected renormalization of the two spin-wave
dispersion parameters $\Delta$ and $D$ \cite{Itoh2016,Jenni2019}.
Here we use the combination of polarized and unpolarized INS experiments with and without polarization analysis to
characterize the spectrum of magnetic excitations in a broader energy range.
At low energies we find a nearly parabolic spin-wave dispersion, and magnon scattering extends to at least 35\,meV, but there is no signature of a Stoner continuum.
However, the magnetic excitations in SrRuO$_3$ are extremely broad.

\section{Experimental}

\sro\ crystallizes in an orthorhombic lattice (space group $Pnma$) at room temperature after undergoing two structural transitions: from cubic to tetragonal at 975\,K and from tetragonal to orthorhombic at 800\,K \cite{Randall1959,Jones1989,Chakoumakos1998}.
This symmetry reduction results in six possible twin-domain orientations which imitate the cubic symmetry.
Therefore, the pseudo-cubic lattice (space group $Pm\bar{3}m$) with lattice parameter $a_{c}=3.93\,\text{\AA}$ is used here and all scattering vectors ${\bf{Q}}$
are given according to this lattice.
The relations between orthorhombic lattice parameters and the cubic directions are as follows: ${\bf{a}} \parallel [1,0,1]_c$, ${\bf{b}} \parallel [{0,1,0}]_c$, and ${\bf{c}} \parallel [{\bar{1},0,1}]_c$ with $a \approx c \approx \sqrt{2}a_{c}$ and $b\approx 2a_{c}$ \cite{Kunkemoeller2017b}.
The sample can be detwinned by applying a magnetic field of more than 1\,T above \tc\ along $[{\bar{1},0,1}]_c$ and then cooling down into the ferromagnetic phase \cite{Kunkemoeller2017b}.
It develops a single domain state where the easy axis (orthorhombic $c$) points along the applied field.
This mono-domain state persists at low temperatures even when the field is turned off \cite{Kunkemoeller2017b}.
Magnetic detwinning was used in experiments at the triple-axis spectrometers PANDA and IN20.

\begin{figure}
\centering
 \includegraphics[width=0.42\columnwidth]{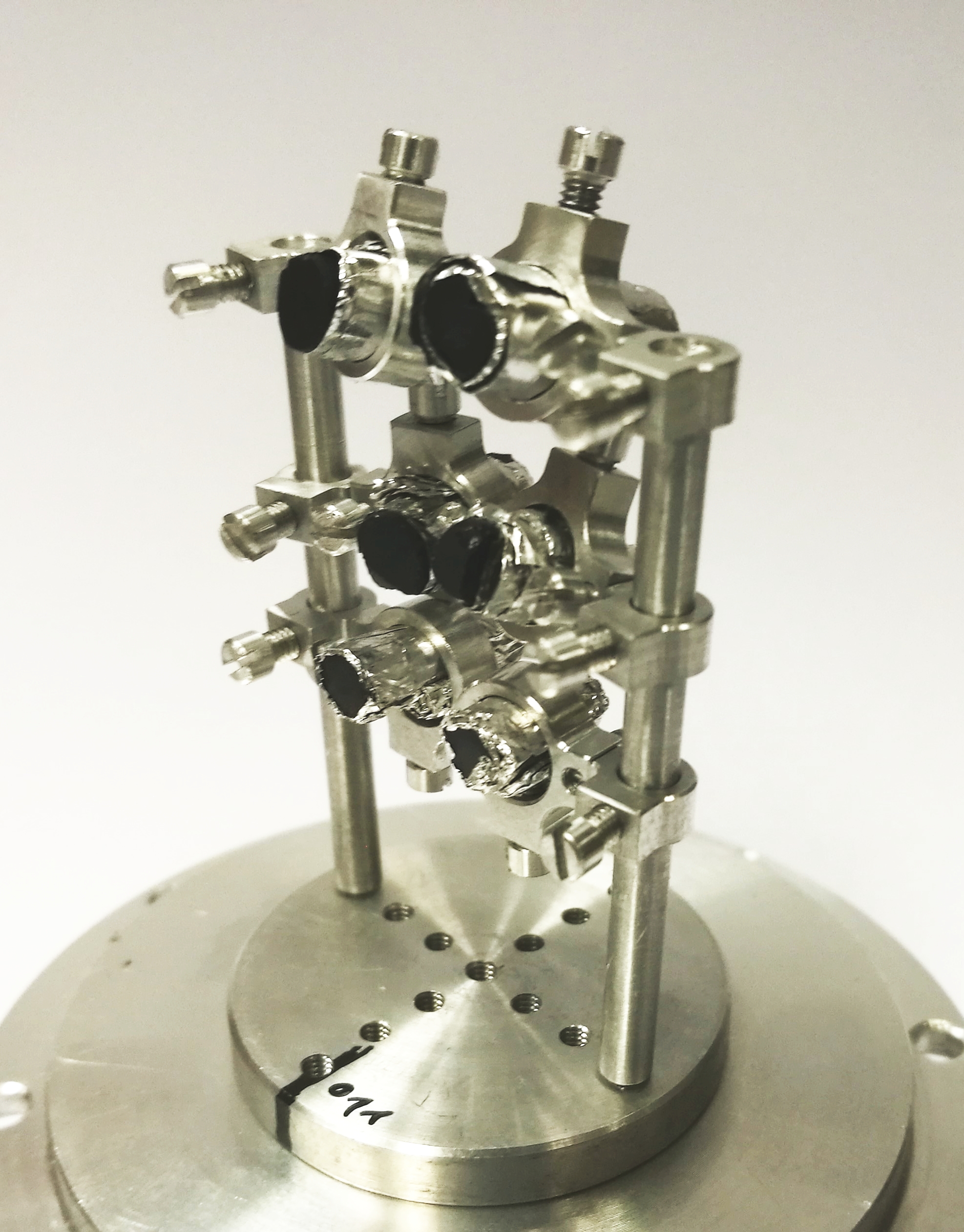}
  \caption{{Coaligned multi-crystal assembly for neutron scattering experiments.} The single crystals of nearly cylindrical shape with a diameter of around 4\,mm and a length of up to 1.5\,cm are individually fixed in aluminium clamps, which are attached to two aluminium rods. This setup enables each crystal to be rotated individually around two axes for easy coalignment.}
  \label{fig:SRO-crystals}
 \end{figure}

The inelastic neutron scattering (INS) data were collected using single crystals grown by the floating-zone method \cite{Kunkemoeller2016}.
The grown crystals exhibit ferromagnetic order below $\text{\tc}=165\,\text{K}$ with a saturation magnetization $\text{M}_\text{sat}=1.6\,\mu_B/\text{f.u.}$ \cite{Kunkemoeller2016}.
The coaligned multi-crystal assembly which was used for most of the neutron scattering experiments is depicted in Fig. \ref{fig:SRO-crystals}, it contains
six crystals with a total mass of about 8\,g.
The compact crystal assembly yields a high material density inside a sample volume of roughly $2\,\text{cm} \times 2\,\text{cm} \times 2\,\text{cm}$.
On PANDA a mounting with only one crystal was used for experiments under magnetic field.

The neutron scattering experiments were conducted at the \tas s 4F and 2T at the Laboratoire L\'{e}on Brillouin (LLB) in Saclay, France, at IN20 at the Institute Laue Langevin (ILL) in Grenoble, France, and at PANDA at the Forschungsneutronenquelle Heinz Maier-Leibnitz (FRM-II) in Garching, Germany.
The time-of-flight data were collected on MERLIN \cite{Bewley2006} at the ISIS Neutron and Muon source  in Didcot, United Kingdom.
The sample was oriented in the $[{1,0,0}]_c/[{0,1,1}]_c$ scattering plane for all scattering experiments.
The \tas s are used with focusing pyrolitic graphite crystals as monochromator and analyzer and scans were performed
with a fixed final energy (values of the final neutron wave vector $k_f$ between 1.5 and 1.57 \AA$^{-1}$ on the cold and 2.662 \AA$^{-1}$ on the thermal instruments, respectively).
Only for the polarized experiment on IN20 we used polarizing Heusler monochromator and analyzer crystals; in this experiment
neutron polarization was guided at the sample by large horizontal field of up to 3.8\,T.
A filter in front of the analyzer (Be filter on PANDA and 4F, pyrolitic graphite filter on 2T) or a velocity selector (on IN20) are used to suppress higher order scattering.
For the 4F and 2T experiments the sample was cooled with a close- cycle refrigerator, while on PANDA and IN20 cryomagnets were used.
On MERLIN the following configurations of incident energy and chopper frequencies were used at 10\,K:
chopper frequency 450\,Hz with incoming energy $E_i$ = 180, 68, 34, and 21\,meV yielding
a resolution at the elastic line of 11, 2.5, 1.2, and 0.6\,meV, respectively; and
chopper frequency 350\,Hz with $E_i$ = 120, 43, and 22meV yielding a  resolution of 7.5, 1.7, 0.7\,meV, respectively.
The sample was rotated by 90\,$^{\circ}$ in 0.5\,$^{\circ}$ steps.
Since the energy resolution improves at higher energy transfer, the resolution is below the bining applied in most cases to
calculate cuts in the four-dimensional data.
At 160\,K only the data with the lower chopper frequency were recorded.
We used the HORACE program suite to calculate the intensity distribution from the data obtained on MERLIN \cite{Ewings2016}.

Data obtained at IN20 and at MERLIN are available at references \onlinecite{data-4-01-1663} and \onlinecite{merlin-data}, repsectively.

\section{Results and Analysis}

\subsection{Unpolarized experiments on triple-axis spectrometers}

\begin{SCfigure*}[\sidecaptionrelwidth][tt]
\centering
 \includegraphics[width=1.4\columnwidth]{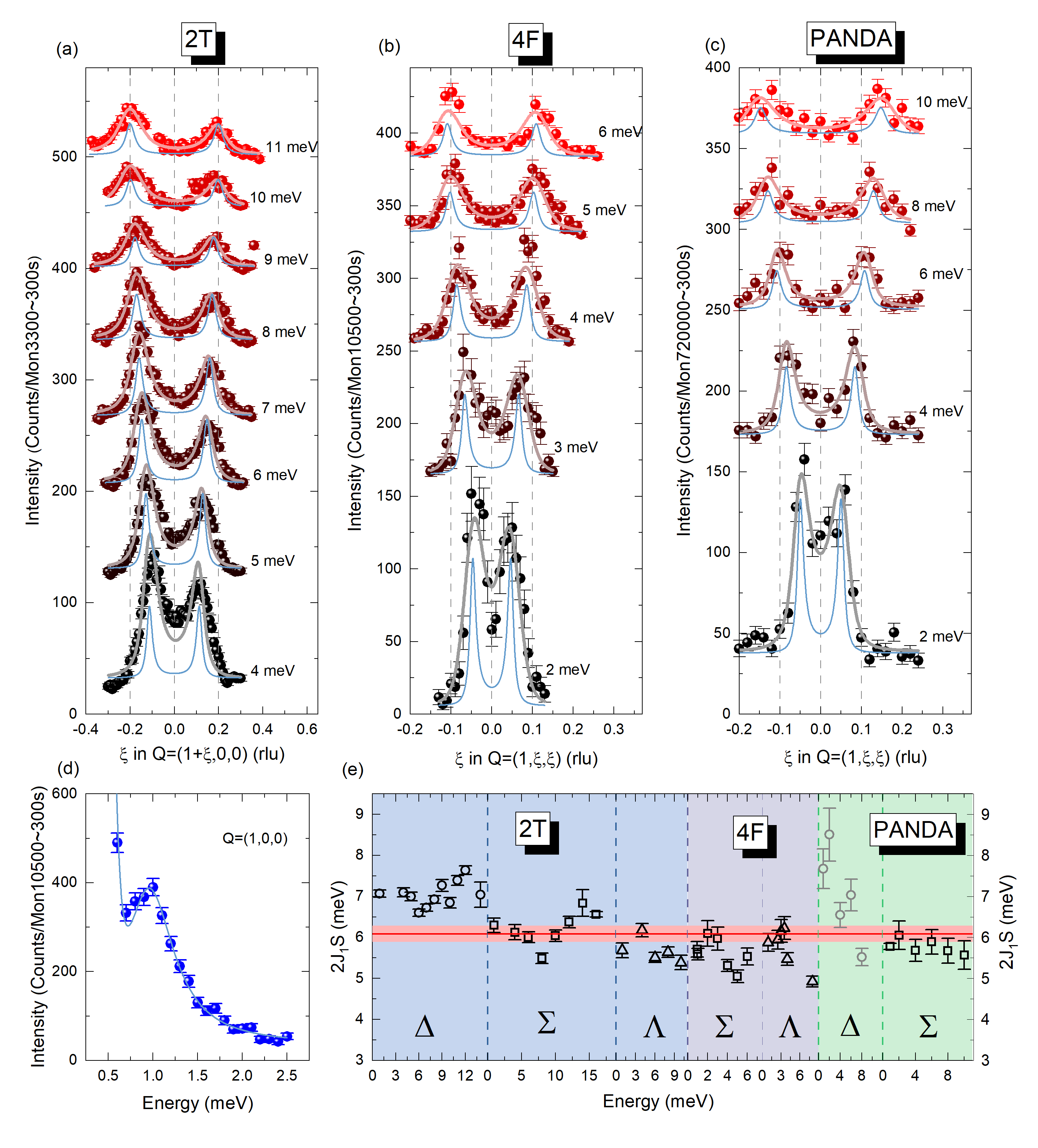}
  \caption[width=0.55\columnwidth]{\label{fig:SRO-TAS-Qscans} \textbf{(a)-(c)} Constant energy scans across the magnon dispersion in \sro\ obtained at T=10\,K on cold triple-axis spectrometers 4F (LLB) and PANDA (MLZ), and on the thermal spectrometer 2T (LLB).
  Note that the PANDA data were measured after detwinning the sample with magnetic field. The magnon scattering was modeled following the ferromagnetic dispersion relation including energy broadening (light blue lines) and then folded with the ${\bf{Q}}$ and $E$ dependent resolution function (colored lines). Data are vertically offset for clarity. \textbf{(d)} Constant ${\bf{Q}}$ scan at ${\bf{Q}}=(1,0,0)$ described with the same dispersion relation showing the anisotropy gap at $\text{T}=10\,\text{K}$ (data taken on 4F). \textbf{(e)} The fitting yields a value of $2J_1S$ for each scan along different high symmetry directions. The data are distinguishable by colored background in respect to the instrument and by symbol shape in respect to the cubic direction (circle: $\Delta$ = [$\xi$,0,0]; square: $\Sigma$ = [0,$\xi$,$\xi$]; triangle: $\Lambda$ = [$\xi$,$\xi$,$\xi$]).
  The weighted average of $2J_1S=6.1(2)\,\text{meV}$ is represented by the red line while the light red area denotes its error margin. Data in gray are not used for the averaging. Data in panels (a) and (b) were already presented in reference \onlinecite{Jenni2019}.}
\end{SCfigure*}

\begin{figure}[htbp]
\centering
 \includegraphics[width=0.65\columnwidth]{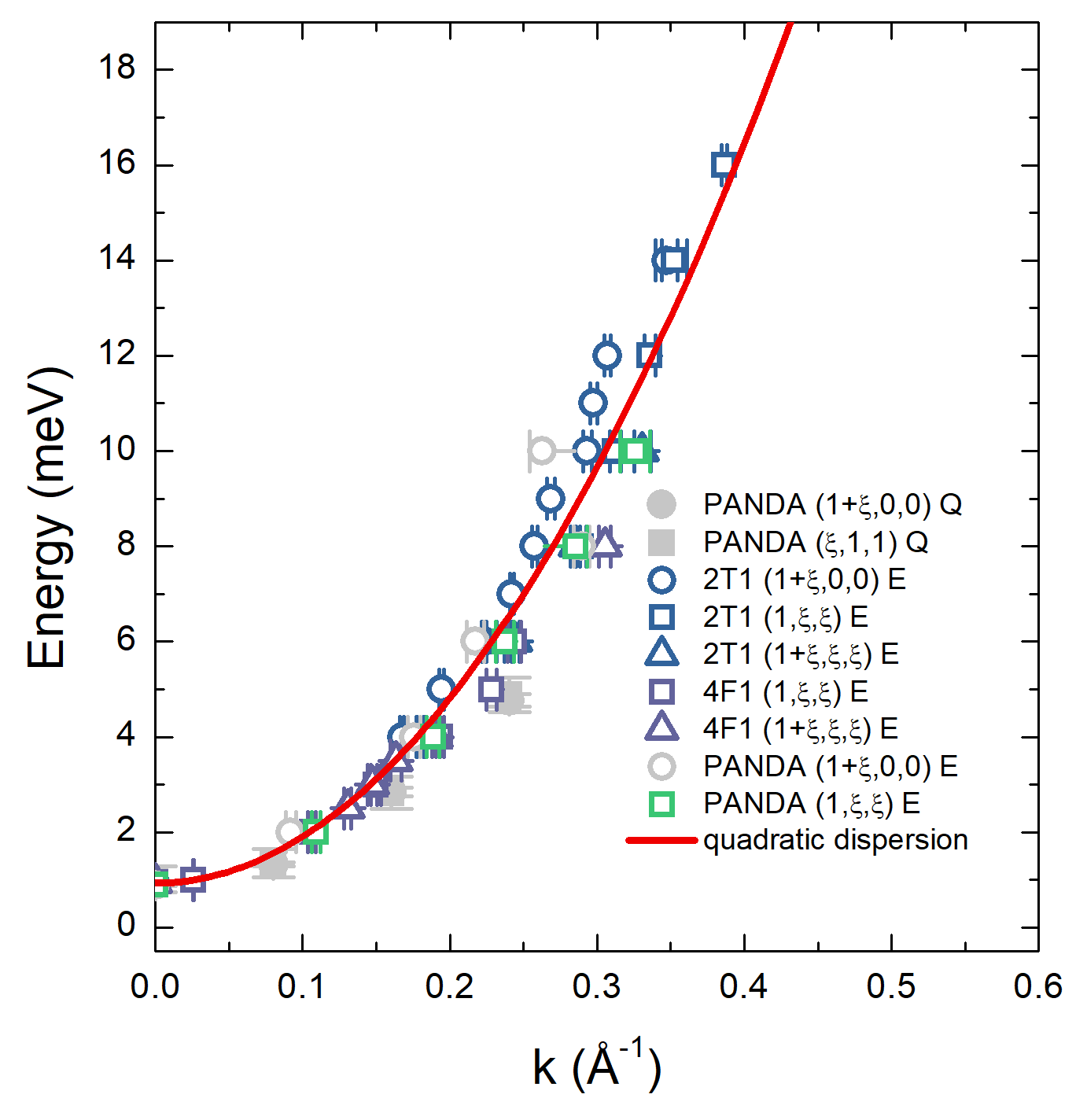}
  \caption{{Low-energy magnon dispersion derived from \tas s.}
   For low energies the magnon energy shows a quadratic dependency on the propagation vector that is given here in absolute units. The $k$ value is calculated from the individually fitted values of $2J_1S$ (given in Fig. \ref{fig:SRO-TAS-Qscans}(e)). The red line represents the parabolic magnon dispersion determined by equation \ref{eqn:quadratic-disp} with the averaged value of $2J_1S$. A similar fit was already presented in reference \onlinecite{Jenni2019}. }
  \label{fig:SRO-TAS-Parabola}
\end{figure}

INS determines the magnon signal in four-dimensional ${\bf{Q}}$-$E$ space,
and the \tas\ allows arbitrarily defined scans.
We combine data taken on instruments installed at cold and thermal neutron moderators to cover a broader energy range.
Typical neutron scattering data from three different spectrometers measuring the magnon signal are displayed in Fig. \ref{fig:SRO-TAS-Qscans}.
Constant energy scans along high-symmetry directions around the ferromagnetic zone center ${\bf{Q}}=(1,0,0)$ reveal the magnon dispersion as the peak position changes with increasing energy transfer (panels (a) to (c) in Fig. \ref{fig:SRO-TAS-Qscans}).
Part of these data were presented in reference \onlinecite{Jenni2019} focussing on the anomalous temperature dependence of the magnon stiffness and anisotropy gap.
Note that the scans cover both sides of the magnetic zone center. Hence, the two peaks appearing in each scan visualize the symmetry of the magnon dispersion.
The cold triple-axis spectrometers 4F1 and PANDA with their high energy resolution enable a direct measurement of the magnon gap via a constant ${\bf{Q}}$ scan at the zone center (Fig. \ref{fig:SRO-TAS-Qscans}(d)).
For the data description the \texttt{MATLAB} based software tool \texttt{Reslib} \cite{Zheludev2007} is used where a given model cross section $\mathcal{S}(\bf{q},E)$ is convoluted with the instrumental resolution function of the specific instrument and fitted to the data.
This procedure enables one to separate the pure excitation-related physics from the effects of the instrumental resolution on the experimental data.
The intensity stemming from the magnon is modeled by a Lorentzian $\mathcal{L}(E)$ with the FWHM $\gamma$ and the amplitude $A$, see equation (\ref{eqn:lorentzian}).
The $E({\bf{q}})$ dependence is modelled by the specified dispersion relation.
To describe the low-energy data, where the tail of the magnetic and nuclear Bragg peak at the zone center yields inelastic scattering, a Gaussian $\mathcal{G}({\bf{q}},E)$ centered at ${\bf{q}}_0=(0,0,0)$ and $E_0=0$ is included in the model cross section.

 The dispersion relation $E(\bf{q})$ can be derived from the general Heisenberg Hamiltonian (equations 1 and 2), although the
 itinerant character of the magnetic order in SrRuO$_3$ strongly limits the applicability of such a model as it will be
 discussed below.

\begin{align}
\cal{H} &= \cal{H}_{SSI} + \cal{H}_{ZFI} + \cal{H}_{EZI} \\
&= -\sum_{{\langle ij\rangle}} 2J_{ij} {\bf{S}}_i \cdot {\bf{S}}_j - \sum_{i} K (S_{i}^{z})^{2} - \mu_{B} g {B} \sum_{i} S_{i}^{z} \label{eqn:HH}
\end{align}

\begin{figure}[htbp]
 \centering
 \includegraphics[width=0.65\columnwidth]{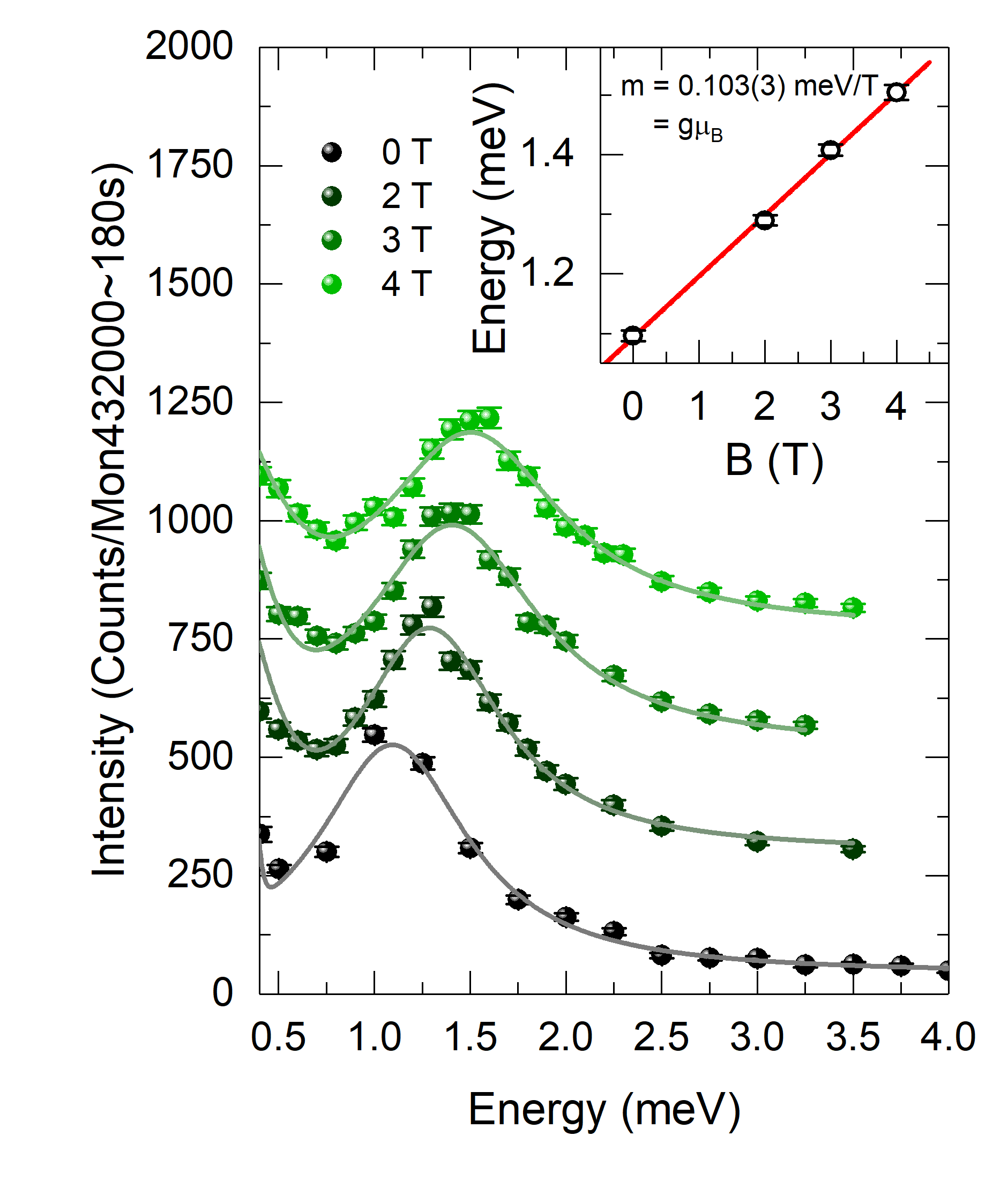}
  \caption{{Magnetic field dependency of the anisotropy gap.} Constant ${\bf{Q}}$ scans at ${\bf{Q}}$ = (0,1,1) are shown for different values of the applied magnetic field (data taken on PANDA). The magnon signal is modeled with a Lorentzian profile combined with a Gaussian background taking the low-energy contribution of the elastic line into account (lines). The data are shifted by a constant offset for better visibility. The inset depicts the magnon gap (Lorentzian peak position) in comparison to the magnetic field. This dependence is fit by a linear function with slope $m$ (red line).}
  \label{fig:SRO-magGap}
 \end{figure}

For the description of spin waves in \sro\ three contributions are considered: (i) the spin-spin interaction \footnote{The sum is built over the pairs $\langle{ij}\rangle$ that appear only once.} with the interaction parameters $J_{ij}$, (ii) the zero-field single-ion anisotropy parameter $K$\footnote{Here an easy-axis anisotropy is assumed.}, and (iii) the electron Zeeman term with the Land\'{e} factor $g$ and the external field ${B}$  that is set parallel
to the magnetization (orthorhombic $c$ or cubic [0,$1$,$\bar{1}$]$_c$).
Note that the indices $i$ and $j$ represent different spins.
For zero external field ${\bf{B}}=0$ and only nearest neighbor interaction $J_1$, the dispersion relation for a ferromagnet with a cubic lattice \cite{Coey2009} is given in equation (\ref{eqn:disp-rel-nn}).
The anisotropy parameter $K$ results in a finite magnon gap $\Delta$ at $\bf{q}=0$.

The magnon dispersion, the Lorentzian distribution and the scattering function are given by:

 \begin{align}
 E_{\bf{q}} &= \Delta + 2J_1S \big[6-2\sum_{i=x,y,z} \text{cos}(2 \pi q_{i})\big] \label{eqn:disp-rel-nn}\\
 {\cal{L}}(E) &= \frac{A}{2\pi} \frac{\gamma}{(E-E_{\bf{q}})^2 + \gamma^2}  \label{eqn:lorentzian}\\
 {\cal{S}}({\bf{q}},E) &= {\cal{G}}({\bf{q}},E) +{\cal{L}}({\bf{q}},E)\cdot(n_E+1),
\end{align}

where $n_E=(exp(E/k_BT)-1)^{-1}$ denotes the Bose population factor.

The constant-${\bf{Q}}$ scan at the zone center (Fig. \ref{fig:SRO-TAS-Qscans}(d)) can be well described with the dispersion model and yields a value of $\Delta$=0.94(3)\,meV at 10\,K for the magnon gap in \sro.
This gap is a manifestation of the single-ion anisotropy of Ru where spin-orbit coupling leads to a preferred alignment of spins along a certain crystallographic direction (easy axis parallel to orthorhombic $c$).
Its size is in agreement with the anisotropy field of $\approx$10\,T determined by magnetization measurements \cite{Kunkemoeller2017b} and with the energy of the ferromagnetic resonance of $\approx250\,\text{GHz}\, \widehat{=}\, 1.03\text{meV}$ observed in time-resolved magneto-optical Kerr effect measurements \cite{Langner2009} as well as in a recent Brillouin light-scattering experiment \cite{Toyoda2022}.
Its temperature dependence is discussed in reference \onlinecite{Jenni2019}.
In the following analysis the magnon gap is fixed in the model cross section, which restricts the fitting of the constant-$E$-scan profiles to background, the amplitude and the single dispersion parameter $2J_1S$.
This parameter contains the coupling constant in the Heisenberg model $J_1$ and is connected with the so called spin stiffness $D$.
The magnon stiffness is defined by approximating the ferromagnetic dispersion relation (\ref{eqn:disp-rel-nn}) for small ${q}$.
The approximation yields a quadratic dispersion relation with the spin stiffness $D$ given in equation (\ref{eqn:quadratic-disp}).

 \begin{eqnarray}
  E_{\bf{k}} \approx \Delta_{mag} + \underbrace{2J_1Sa_{c}^2}_{D} k^2
 \label{eqn:quadratic-disp}
 \end{eqnarray}

Here, $k=q\frac{2\pi}{a}$ denotes the magnon propagation vector in absolute units. The theoretical description of the experimental data is displayed by the lines in Fig. \ref{fig:SRO-TAS-Qscans}(a)-(d).
The light blue lines represent the cross section model for the $\bf{q}$ and $E$ values specific to each scan and the lines in same colored as the data visualize the convolution of the model with instrumental resolution.
By comparison of the model with the convolution the influence of the instrumental resolution on the experimental data becomes visible.
The ellipsoid shape of the resolution  function in ${\bf{Q}}$-$E$ space leads to a focusing effect where the magnon signal on one side of the constant $E$ scan is enhanced.
Additionally the instrumental function governs the width of the magnon peaks in the experimental data since it significantly broadens the peaks derived from the model.
Nevertheless the description of experimental data clearly needs a finite width of the model which indicates that the magnons are intrinsically broadened, even at 10\,K.  The intrinsic width of the model is set to 40\,\% of the specific energy.
The twinning in the \sro\ crystals can imply some broadening, since this superposes different directions of the orthorhombic lattice.
The pseudo-cubic direction $[1,0,0]_c$ is parallel to the long orthorhombic axis $b$ (in $Pnma$) of one twin orientation and parallel to the in-plane diagonals $[1,0,\pm1]_o$ for the other orientations \cite{Kunkemoeller2017b}.
At low temperatures the splitting of the orthorhombic lattice constants renormalized to the pseudocubic ones amounts to less than 0.6\% \cite{Kiyama1996}.
Therefore, the twinning related superposition of scattering vectors does not account for the large effects observed unless the magnon dispersion becomes very
anisotropic, which appears unlikely.
The sizable intrinsic broadening of the magnon results most likely from the coupling to electron-hole excitations called Landau damping \cite{Buczek2011,Savrasov1998} in agreement with the kink of electric resistivity at the ferromagnetic transition\cite{Allen1996,Klein1996}. In addition non-linear scattering with spin excitations can limit the lifetimes
of the magnons \cite{Solontsov1995,Solontsov2005}.
The temperature dependent INS data in reference \onlinecite{Jenni2019} and the discussion below indicate further enhanced broadening at higher temperatures.

As mentioned before, the peak positions in the constant $E$ scans are determined by the $2J_1S$ parameter which is fitted in the analysis of each scan.
Fig. \ref{fig:SRO-TAS-Qscans}(e) shows the resulting $2J_1S$ values for each scan separated for the three instruments (background color) and the different high symmetry directions (symbol).
Unfortunately the fitting of all scans in a multi-fit routine using one set of generalized fitting parameters is not feasible with the used software tool due to the individual backgrounds of the different instruments.
Therefore the results of fitting all scans are averaged and yield a general $2J_1S$ of 6.1(2)\,meV.
This translates to a spin stiffness in \sro\ of $D=94.2\pm3.0\,\text{meV\AA}^2$.
This averaged spin stiffness describes the low energy data of all instruments reasonably well as one can see in the $E$-$k$ dependency for the magnon signal in Fig. \ref{fig:SRO-TAS-Parabola}.
Here the $k$ values are calculated from the fitted $2J_1S$ of each individual scan [given in Fig. \ref{fig:SRO-TAS-Qscans}(e)] following the quadratic relation (\ref{eqn:quadratic-disp}) and plotted against the energy.
This differs from the determination of spin stiffness $D$ in reference  \onlinecite{Jenni2019} where D results from the approximated quadratic dispersion model for small $q$ which is fitted to the $E$-$k$ data extracted from the constant energy scans.
The coupling constant $J_1$ can be estimated to 3.8(1)\,meV by determining the spin $S$=0.8 from the saturation magnetization of 1.6\,$\mu_B/\text{Ru}$ \cite{Kunkemoeller2017b} with a $g$ factor of 2 \cite{note-g}. This value is close to the expected spin for a low-spin state stabilized through the
strong splitting of $t_{2g}$ and $e_g$ orbitals \cite{Khomskii2014}.

A possibility to directly measure the $g$ factor is the magnetic field dependency of the magnon gap.
Applying an external magnetic field adds a Zeeman term in the Hamiltonian and a constant shift of the magnon dispersion: (\ref{eqn:HH}).
We therefore have to extend the dispersion relation (\ref{eqn:disp-rel-nn}):

  \begin{eqnarray}
  E_{\bf{q}} = \Delta + 2J_1S \Big[6-2\sum_{i=x,y,z} \text{cos}(2 \pi q_{i})\Big] + g\mu_{B}B.
 \label{disp-rel-zeeman}
  \end{eqnarray}


The magnon energy at the zone center ${\bf q}=0$ increases linearly with the magnetic field $B$ applied along orthorhombic $c$ or cubic [0,$1$,$\bar{1}$]$_c$.
The magnetic field dependency of the magnon signal was studied on PANDA with a single sample crystal.
Figure \ref{fig:SRO-magGap} displays the constant ${\bf{Q}}$ scans measured at the ferromagnetic zone center ${\bf{Q}}=(0,1,1)$ indicating that
the magnon gap increases with increasing field.
The gap value determined by the Lorentzian peak position of the fit indeed exhibits a linear correlation to the external field (see inset of Fig. \ref{fig:SRO-magGap}).
The slope $m$ of the linear fit is equal to $g\mu_{B}$ yielding a value of $g$=1.78(5) \cite{note-g}.

Usually the $g$ factor in $4d$ transition-metal oxides with a high crystal-field splitting like in \sro\ is assumed to consist mainly of the spin contribution $g_S=2$
because the orbital moment is quenched.
However, sizable \soc\ can partly recover the orbital moment yielding a $g$ factor deviating from 2 \cite{Khomskii2014}.
The orbital moment of \sro\ is found by x-ray magnetic circular dichroism to be very small \cite{Okamoto2007,Agrestini2015}.
Okamoto \etal\ reported an orbital moment of $0.04(4)\,\mu_B$ \cite{Okamoto2007}, and Agrestini \etal\ determined $L_z/2S_z$ ratios of 0.01 with $L_z = 0.01(1)\,\mu_B$ \cite{Agrestini2015}.
These experimental reports are supported by DFT calculations which obtain an orbital moment three orders of magnitude smaller than the spin moment \cite{Kunkemoeller2019}.
Our INS determination of the $g$ factor is consistent with a small but finite orbital moment.

The strength of the magnetic interaction parameters is of the same order as those in insulating Ca$_2$RuO$_4$ \cite{Kunkemoeller2017}.
Anisotropic magnetic interaction parameters in SrRuO$_3$ were determined by density functional theory but the agreement with the experimental stiffness and with the magnon gap is poor \cite{Ahn2022}. The latter calculations determines also the Dzyaloshinski-Moriya interaction, which, however, has very little impact on the magnon dispersion. 
A precise measurement of the canting angle of the ferromagnetic moments in SrRuO$_3$ is better suited to experimentally determine this interaction. 

\begin{figure}[!t]
 \centering
 \includegraphics[width=1.0\columnwidth]{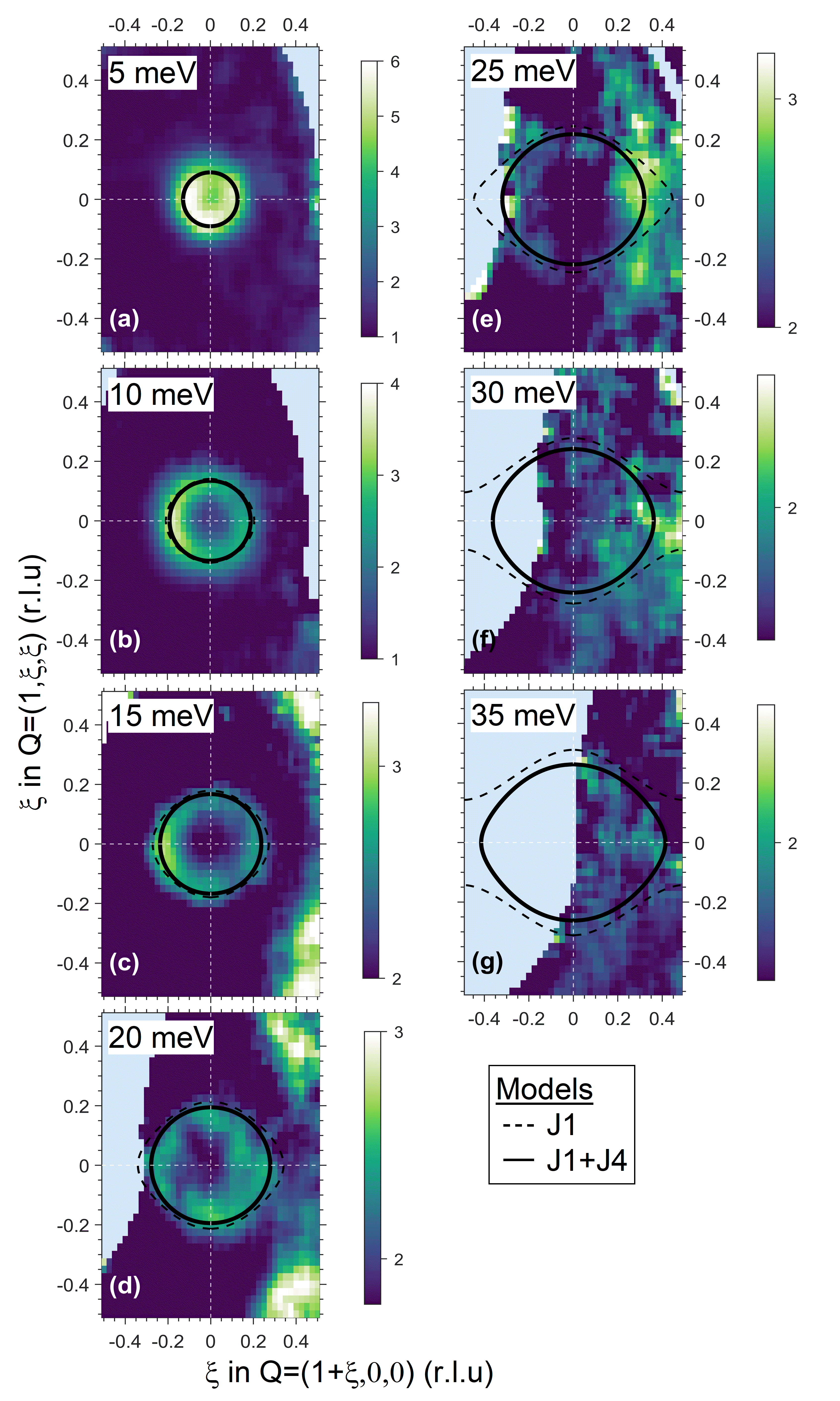}
  \caption{{Magnon dispersion measured by time-of-flight technique.} Constant $E$ cuts at 10\,K of the [$\xi$,0,0]/[0,$\xi$,$\xi$] plane for different energies display the ring shape of the magnon signal. The diameter of the ring increases with energy indicating the magnon dispersion.  The axis length ratio of 1:$\sqrt{2}$ reflects the geometrical factor. The low-energy data at 5 and 10\,meV (panels (a) and (b)) are taken with the incident energy of 22\,meV while the data at 15 and 20\,meV ((c),(d)) are taken with the incident energy of 43\,meV and the data at 25, 30, and 35\,meV ((e)-(g)) are taken with the incident energy of 68\,meV.
  The used integration limits for these maps are $-0.1 \leq \eta \leq 0.1$ in $[1,-\eta,\eta]$ and $E \pm 2.5\,\text{meV}$.
  The time-of-flight data are overlaid with the ferromagnetic dispersion model with only nearest-neighbor interaction $J_1$ taken from the analysis of the \tas\ data (dashed line) and with the combination of nearest-neighbor and $J_4$ interaction (black lines).}
  \label{fig:SRO-TOF-Donuts-lowT}
 \end{figure}

Time-resolved magneto-optical Kerr effect measurements on \sro\ thin films also quantify the linear field dependence of the ferromagnetic resonance \cite{Langner2009}.
They report a slope of $\approx 17\,\frac{\text{GHz}}{\text{T}}$ which corresponds to $0.07\,\frac{\text{meV}}{\text{T}}$ and thus an even smaller $g$ factor of 1.21.
The magnetization of SrRuO$_3$ amounts to $\mu_0M$=0.31\,T and thus demagnetization effects cannot explain such a large deviation at high magnetic fields.
In our experiment on PANDA the zero-field result was measured after cooling the sample in a strong field yielding a similar macroscopic magnetization as at high field
so that the demagnetization corrections are roughly the same at all fields.
The much slower slope of the optical experiment \cite{Langner2009} must stem from the fact that the external field
is not applied parallel to the easy axis of the ferromagnetic phase.
Therefore, the external field at least partially acts against the local anisotropy, and the slope of the resonance does not correspond to $g\mu_{B}$.

\subsection{Unpolarized experiments on the time-of-flight spectrometer Merlin}

\begin{SCfigure*}[\sidecaptionrelwidth][tt]
\centering
 \includegraphics[width=1.35\columnwidth]{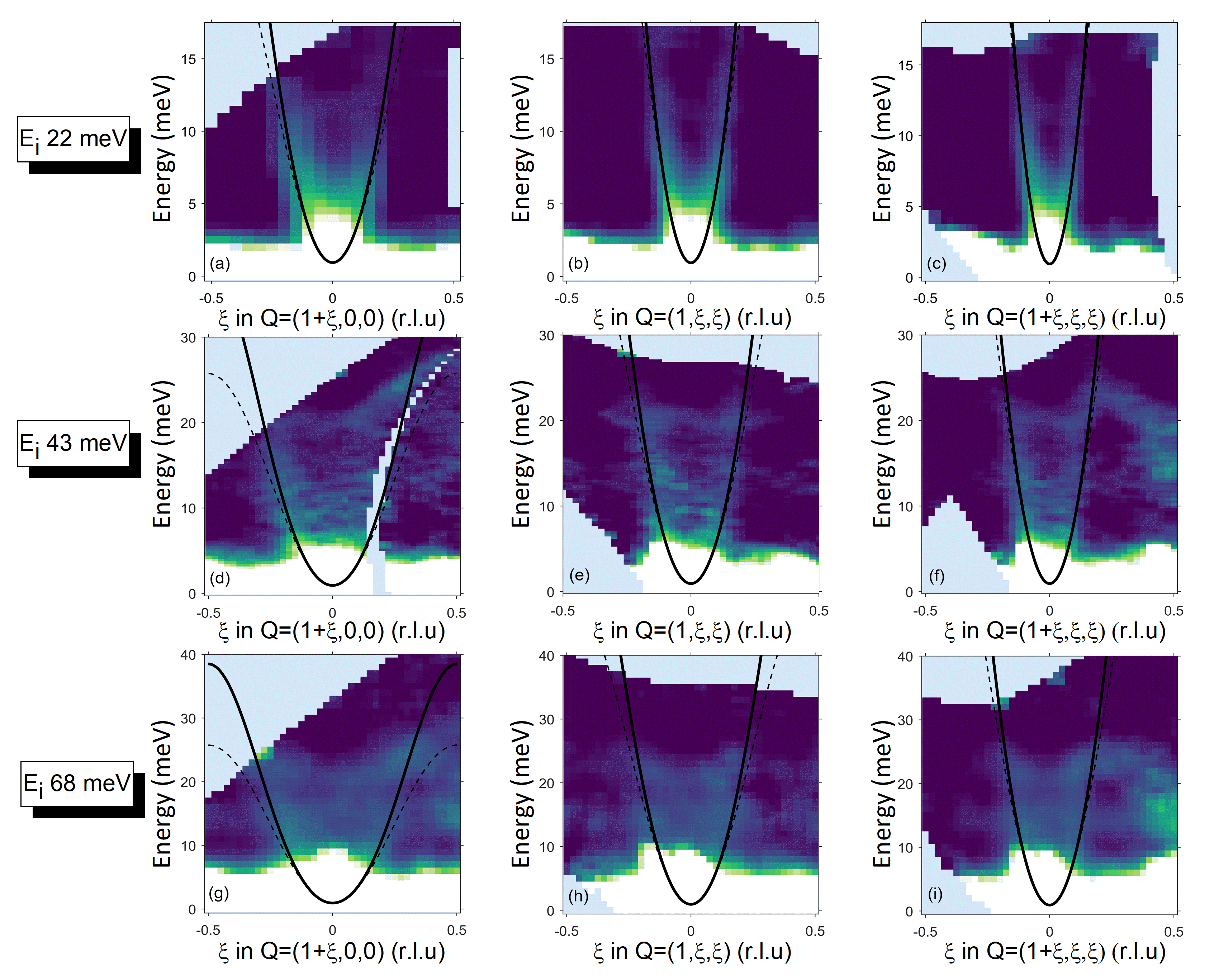}
  \caption[width=0.6\columnwidth]{{Magnon dispersion measured by time-of-flight technique.} Constant ${\bf{Q}}$ cuts at 10\,K along the cubic high-symmetry directions around ${\bf{Q}}$ = (1,0,0) show the magnon dispersion. The panels are sorted in rows where each row represents the data of a certain incident energy. From the top to the bottom the data are taken with 22\,meV (panels (a)-(c)), 43\,meV (panels (d)-(f)), and 68\,meV (panels (g)-(i)), respectively. The intensity range (colorbar) is set identical in the panels of each row for better comparability.
  The integration limits are $-0.1 \leq \eta \leq 0.1$ in $[1,-\eta,\eta]$ and $-0.1 \leq \zeta \leq 0.1$ in the direction perpendicular to the respective high-symmetry direction.
  The time-of-flight data are overlaid with the ferromagnetic dispersion model with nearest-neighbor interaction $J_1$ (dashed line) and with the combination of nearest-neighbor and $J_4$ interaction  (black lines). }
  \label{fig:SRO-TOF-QE-lowT}
 \end{SCfigure*}

The investigation of the magnon dispersion using \tas s becomes increasingly difficult for high energies, where
phonon contributions and spurious signals appear.
The time-of-flight technique can deliver a complete picture of the Brillouin zone with its different excitations.
It uses higher initial energies than the  \tas s, which enhances the access in ${\bf{Q}}$-$E$ space to lower $Q$ values and thus
favors the observation of magnetic signals due to the form factor.
The magnon dispersion of \sro\  was studied using the time-of-flight spectrometer Merlin at the ISIS Neutron and Muon Source.
The \tof\ technique enables one to collect data simultaneously for several incident energies $E_i$. This creates comparable data sets with different energy resolution and range.
The presented data are taken from the data sets with an incident energy of 22, 43, and 68\,meV since they yield the clearest picture of the magnon signal.

\begin{figure}[htbp]
 \includegraphics[width=0.98\columnwidth]{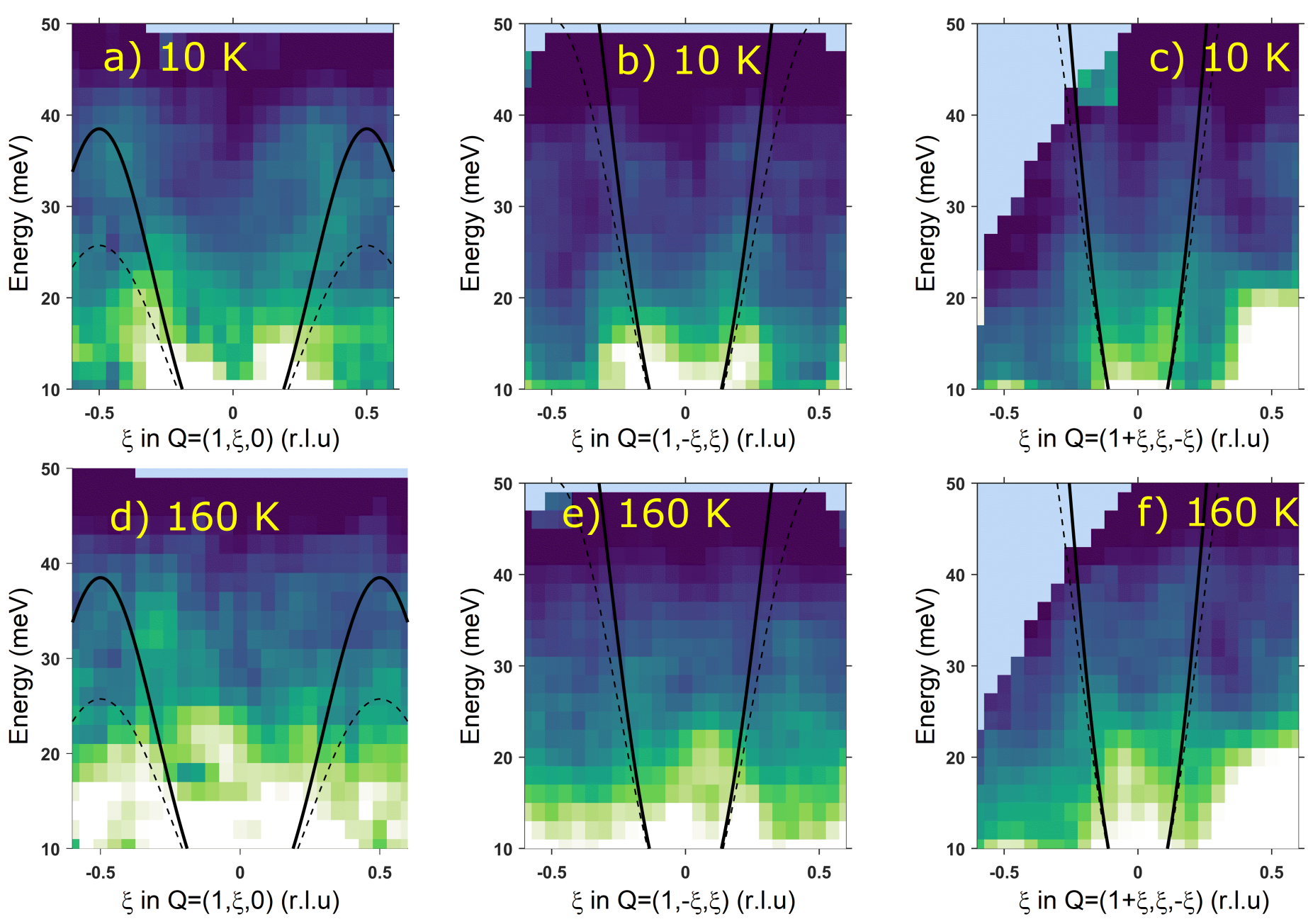}
  \caption{{ } High-energy data taken on the time-of-flight spectrometer Merlin with an incoming energy of 120\,meV. The upper panels
  present the intensity maps of energy versus ${\bf Q}$ vector obtained at 10\,K and the lower panels the data taken at 160\,K.  }
  \label{fig:120meV}
 \end{figure}

The magnon dispersion can be visualized by two-dimensional cuts through the four-dimensional ${\bf{Q}}$-$E$ space.
Two different representations of the magnon dispersion are used: (i) constant $E$ cuts of the scattering plane $[\xi,0,0]/[0,\xi,\xi]$ which represent horizontal cuts through the $q$-$E$ dispersion parabola (see Fig. \ref{fig:SRO-TOF-Donuts-lowT}) and (ii) constant ${\bf{Q}}$ cuts which display the $q$-$E$ dependency of the magnon signal along one of the the cubic high-symmetry directions (see Fig. \ref{fig:SRO-TOF-QE-lowT} and \ref{fig:120meV}).
The data are integrated in the vertical direction $[0,\xi,-\xi]$ by $\pm0.1\,\text{r.l.u}$ and in energy by $\pm2\,\text{meV}$.
To optimize the presentation of the magnon dispersion the different energies are taken from different incident energies $E_i$.
Fig. \ref{fig:SRO-TOF-Donuts-lowT}(a) and (b) result from the with $E_i = 22\,\text{meV}$, (c) and (d) are taken from the data with $E_i = 43\,\text{meV}$ and panels (e)-(g) display the data with $E_i = 68\,\text{meV}$.
For these maps around the magnetic Bragg peaks symmetrization yields a slightly better statistics, while
the variation of the background strongly affects symmetrization of $q$-$E$ maps so that we refrained from it.
The branches of the magnon dispersion are not clearly visible in the data with the highest incident energy $E_i = 180\,\text{meV}$, and also the
$E_i = 180\,\text{meV}$ are only useful at higher energy transfer.
In the two-dimensional constant $E$ cuts the magnon exhibits a ring shape whose radius is increasing with increasing energy indicating the dispersion of the magnon.
Fig. \ref{fig:SRO-TOF-QE-lowT} displays the two-dimensional $Q$-$E$ cuts along the three cubic high-symmetry directions where the dispersion parabola of the magnon becomes visible. The data taken with $E_i = 120\,\text{meV}$ extends to higher energy transfer and a zoom on this high-energy range is shown in Fig. \ref{fig:120meV}.
Note that the data are always integrated by $\pm0.1\,\text{r.l.u}$ in the two corresponding perpendicular directions.
The low-energy part is dominated by the tail of the elastic scattering, which is visible by the bright area for all $\xi$.
The expansion of the elastic scattering into the inelastic regime depends on the energy resolution and increases therefore with the incident energy.
The magnon signal is clearly visible as a parabola in its $q$-$E$ dependency.
Phonon contributions and how they disperse can be seen for example in Fig. \ref{fig:SRO-TOF-QE-lowT}(f),(i) at ${\bf{Q}}=(1.5,0.5,0.5)$ around $E=15\,\text{meV}$.
The data with higher incoming energy shows that there are two phonon modes in this energy range at 11.5 and 19\,meV.
All ${\bf{Q}}$ versus $E$ intensity maps show a flat intensity at 20meV that, however, can be safely attributed to a phonon as it is observed with enhanced intensity at $\bf{Q}$=(2,1,0) and (3,0,0) and as it is found in the non-spin-flip channel in the polarized experiment performed on the IN20 spectrometer, see below.
Only in some of the plots one also sees a weak signal at $\bf{Q}$=(1,0,0) at 12\,meV, which
however seems to stem from a phonon branch with essentially flat dispersion along (1,$\xi$,$\xi$) which can leak to (1,0,0) due to resolution and integration effects.

The data in Fig. \ref{fig:SRO-TOF-QE-lowT} suffer from heavy phonon contaminations around ${\bf{Q}}=(1.5,0,0)$ which can be easily mistaken as the magnon signal.
This phonon contamination is also clearly visible in the in-plane scattering where it appears as intense scattering at the zone corners ${\bf{Q}}=(1.5,0.5,0.5)$ and ${\bf{Q}}=(1.5,-0.5,-0.5)$ for $E=15\,\text{meV}$ (Fig. \ref{fig:SRO-TOF-Donuts-lowT}(c)).
It disperses inwards and is visible as a strong broad signal at ${\bf{Q}}=(1.5,0,0)$ and $E=25\,\text{meV}$ (Fig. \ref{fig:SRO-TOF-Donuts-lowT}(e)).
The phonon dispersion study for \sro\ indeed reveals a phonon at the equivalent position ${\bf{Q}}=(2.5,0,0)$ and the energy $E=25\,\text{meV}$ \cite{Jenni-unpublished}.

 \begin{figure}[htbp]
 \includegraphics[width=0.9997\columnwidth]{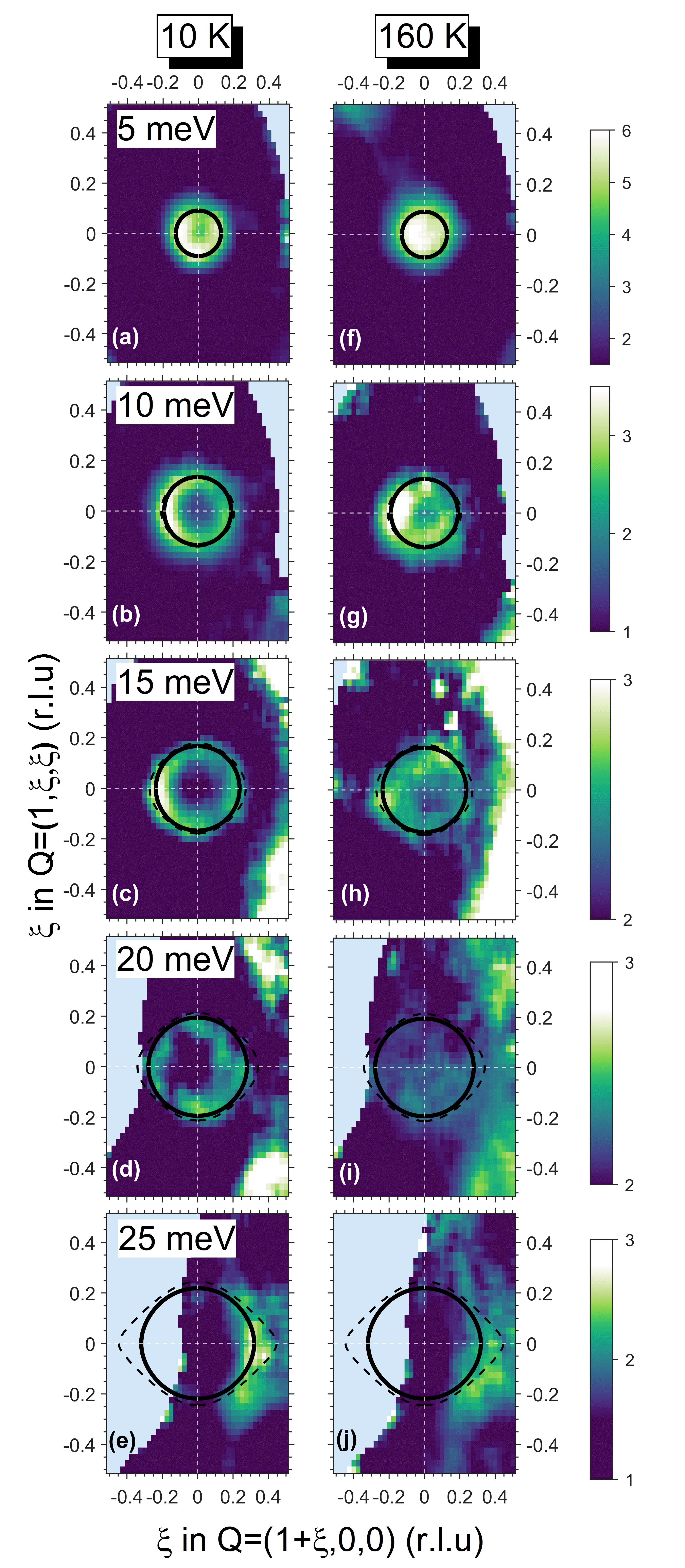}
  \caption{Constant $E$ cuts of the [$\xi$,0,0]/[0,$\xi$,$\xi$] plane display the ring shape of the magnon signal.
 The magnon signal in the ferromagnetic phase at 10\,K ((a)-(e)) is compared with the data at 160\,K ((f)-(g)).
 The intensity range (colorbar) is the same for each energy, and the axis length ratio of 1:$\sqrt{2}$ reflects the geometrical factor.
 The low-energy data at 5 and 10\,meV are taken with the incident energy of 22\,meV while the higher-energy data are taken with $E_i$=43\,meV.
 Data are overlaid with the magnon dispersion calculated with only $J_1$ (dashed lines) and with the combination of $J_1$ and $J_4$ (black lines). }
  \label{fig:SRO-TOF-QQ}
 \end{figure}

To compare the results of the time-of-flight measurement with the triple-axis spectrometer results the theoretical dispersion according to the Heisenberg model of a ferromagnet is overlaid on the experimental data.
Firstly the model in equation \ref{eqn:disp-rel-nn} with only nearest-neighbor coupling $2J_1S$ and the anisotropy gap $\Delta_{mag}$ determined by the triple-axis experiments is compared with the time-of-flight data. It is obvious that this model (black dashed line in Figures \ref{fig:SRO-TOF-Donuts-lowT} and \ref{fig:SRO-TOF-QE-lowT}) only describes the low energy-part of the dispersion.
Note that the triple-axis spectrometer data only cover energies below 16\,meV.
In general the simple nearest-neighbor model underestimates the magnon stiffness at high energy as the experimental parabolas become
tighter and the rings smaller than what is expected with this most simple model.

The underestimation of the higher magnon energies is best seen in the  [$\xi$,0,0] direction, see Fig. \ref{fig:SRO-TOF-Donuts-lowT}, \ref{fig:SRO-TOF-QE-lowT}, and \ref{fig:120meV}.
In order to obtain a better description we add interaction parameters to further distant neighbors in the primitive cubic lattice.
Since the magnon stiffness of the quadratic dispersion is very well determined by the triple-axis experiments we kept this value fixed.
Adding the next-nearest-neighbor interaction $J_2$ between two Ru ions at $\sqrt{2}a$ however does not modify the [$\xi$,0,0] dispersion under the constraint of constant stiffness.
The same holds for the next-next-nearest shell, $J_3$, at a distance $\sqrt{3}a$.
Only with a negative (antiferromagnetic) value of the fourth-neighbor interaction $J_4$ between Ru ions at a distance of $2a$ we can model the steepening of the [$\xi$,0,0] dispersion at higher energy.
The dispersion of this $J_1$-$J_4$ model is given in equation \ref{eqn:disp-rel-nnn}.

\begin{equation}
\begin{aligned}
  E_{\bf{q}} = \Delta_{mag} & + 2J_{1}S \Big[6-2\sum_{i} \text{cos}(2 \pi q_{i})\Big] \\
  & + 2J_{4}S \Big[6-2\sum_{i} \text{cos}(4 \pi q_{i})\Big]
 \label{eqn:disp-rel-nnn}
\end{aligned}
  \end{equation}

The coupling term $2J_4S$ is determined by fitting the ($\xi$,0,0) values extracted from the constant energy cuts with the constraint
of fixed magnon stiffness.
The best agreement is achieved for the values given in Tab. 1.
The modified model also better describes the parabolas in Fig. \ref{fig:SRO-TOF-QE-lowT} and \ref{fig:120meV} although the difference between the models is small for the displayed energy region in [0,$\xi$,$\xi$] and [$\xi$,$\xi$,$\xi$] direction.

While the additional parameter yields a qualitative description of the stiffening of the dispersion at higher energies, it
appears more likely that the physical mechanism for this effect is different. The itinerant character of the magnon dispersion
limits the applicability of the model of local-moment interactions. Unfortunately the data quality is too limited at
higher energies for a deeper analysis.

 \begin{SCfigure*}[\sidecaptionrelwidth][tt]
\centering
\includegraphics[width=1.35\columnwidth]{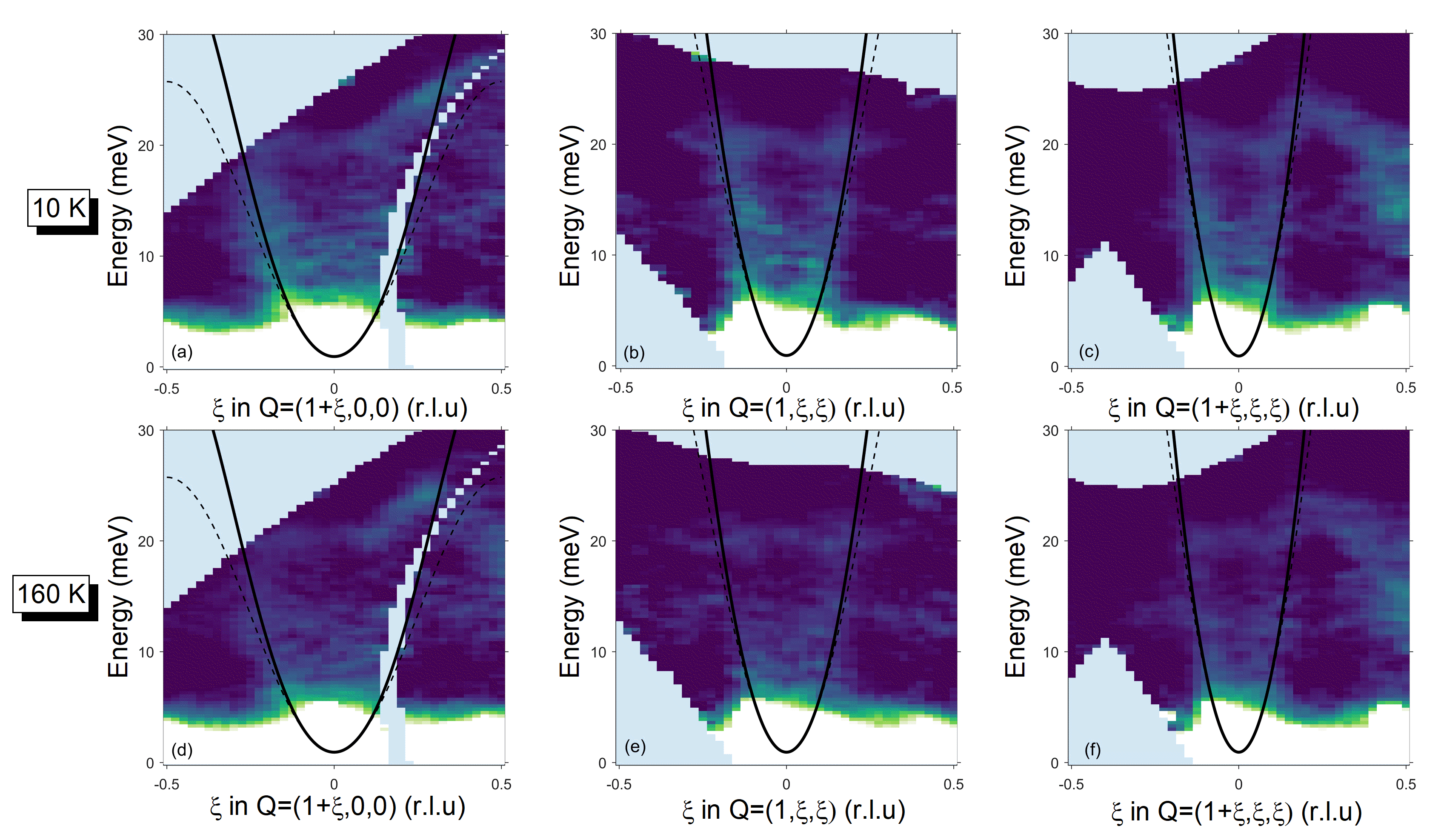}
  \caption[width=0.65\columnwidth]{Energy versus ${\bf{Q}}$ maps along the cubic high-symmetry directions around ${\bf{Q}}$ = (1,0,0) at different temperatures. The data are taken with the incident energy of 43\,meV. The top row ((a)-(c)) displays the low-temperature data at 10\,K while the bottom row ((d)-(f)) shows the data around the phase transition at 160\,K. The intensity range (colorbar) is in all panels the same for better comparability. Data are overlaid with the ferromagnetic dispersion calculated with nearest-neighbor interaction (dashed line) and with the combination of nearest-neighbor and $J_4$ interaction (black lines).}
  \label{fig:SRO-TOF-QE}
 \end{SCfigure*}

   \begin{table}[h]
 \centering
 \caption{\label{tab:sdpara}{Magnetic model parameters of \sro\ determined by inelastic neutron scattering}
 The isolated interaction parameters are calculated from the values $2J_iS$ determined by
 the magnon dispersion by assuming $S$=0.8. $J_i'$ denote the values of the model with nearest and fourth-nearest interaction that keeps
 the magnon stiffness unchanged. }
 \begin{ruledtabular}
 \begin{tabular}{cccccc}
 \hline
  $\Delta_{mag}$  & $D$                   & $2J_1S$   & $J_1$   & $J_1'$  & $J_4'$  \\
  $[$meV$]$       &   $[$meV\AA$^2$$]$ &   $[$meV$]$   &   $[$meV$]$ &  $[$meV$]$  &  $ [$meV$]$ \\
 \hline
 0.94(3)         & 94(2)                 & 6.1(2)  & 3.8(1) & 5.9  & -0.5 \\
 \hline
\end{tabular}
\end{ruledtabular}
 \end{table}

In a metallic ferromagnet the spin-wave dispersion following the Heisenberg model of localized moments is cut off at a finite energy above which magnetic excitations become electron-hole pair excitations between bands of opposite spin, the so-called Stoner continuum \cite{Blundell2001}.
The occurrence of these Stoner excitations in ${\bf{Q}}$-$E$ space can be complex since the band structure in SrRuO$_3$ has multiple bands with changing band splittings throughout the Brillouin zone.
The signature of Stoner excitations in neutron scattering is a broadening of the spin-wave excitations while their intensity decreases rapidly for increasing energy as they enter the continuum \cite{Mook1973}.
Indeed, especially in the data taken with $E_i=68\,\text{meV}$, the intense magnon scattering seems to be reduced above 25\,meV [see Fig. \ref{fig:SRO-TOF-QE-lowT}(h),(i)].
Above this energy some magnon scattering persists but its intensity is significantly lower.
The same behavior is seen in Fig. \ref{fig:SRO-TOF-Donuts-lowT}(f) and (g), where the ring shaped magnetic scattering is significantly lower at 30\,meV and also seems to be broadened.
Nevertheless the scattering is still structured as the ring shape is clearly visible.
For the analysis of the high-energy range, the data taken with the incoming energy of 120\,meV are informative, see Fig. \ref{fig:120meV}.
The extension of the magnon dispersion up to at least $\sim$35\,meV is unambiguous although the signal remains weak.
These high-energy data also confirm the steepening of the dispersion compared to a simple next-neighbor Heisenberg model.
The time-of-flight data of three dimensional material like \sro\ suffer from the fact that it is not possible to fully integrate over one dimension as it is done for example in two-dimensional layered materials.
Nevertheless it is possible to identify spin-wave excitations up to an energy of $\sim$35\,meV.

The high-energy suppression of the magnon signal strength and the pronounced broadening strongly disagree with the simple 
local-moment picture and underline the itinerant character of the ferromagnetic order in SrRuO$_3$. 
In the elementary and other simple ferromagnets similar suppression of intensity and
reduced magnon lifetimes were observed in experiment \cite{Mook1985,Yethiraj1991,Brookes2020,Moriya1985} and in density-functional
theory calculations \cite{Lowde1970,Moriya1985,Savrasov1998,Buczek2011}. The interaction with the Stoner continuum, which can be
rather complex in a multiorbital system like SrRuO$_3$, causes strong Landau damping and impacts the intensity. 
Similar effects are also discussed in iron- or copper-based superconductors \cite{Dai2015,Pailhes2004}, but spin fluctuations
posses an antiferromagnetic character in these materials.  

Beside the analysis of the low-temperature magnon dispersion in the ferromagnetic phase we also studied the magnon dispersion close to the phase transition at 160\,K in the Merlin experiments.
In figures \ref{fig:120meV}, \ref{fig:SRO-TOF-QQ}, and \ref{fig:SRO-TOF-QE} the high-temperature data are directly compared to the low temperature in form of the same two-dimensional cuts of the ${\bf{Q}}$-$E$ space.
Note that for the incident energy of 68\,meV no high-temperature data were measured.
As the intensity ranges represented by the color bars are scaled equally for both temperatures it becomes evident that the magnon signal increases with temperature.
Inside the ferromagnetic phase the increase of magnetic scattering intensity can be explained by the Bose factor.
However at 160\,K, close to the phase transition, the magnetic scattering should decrease faster than the Bose factor enhancement since the magnetization, i.e. the ordered moment, decreases.
Also the 8\,meV constant-$E$ scans presented in reference \cite{Jenni2019} clearly shows the persistence of inelastic magnetic correlations well above the Curie temperature.
The strong magnetic scattering at the phase transition and at higher temperature can be explained as paramagnon scattering which
follows the same dispersion as the long-range order excitations at 10\,K.
To analyze the change of the magnon dispersion the dispersion model derived from the low-temperature data are also plotted in the high-temperature data.
There is no indication that the magnon stiffness changes significantly as the model still describes the intensity maxima.
This supports the magnon softening at low temperatures and the detailed temperature dependence reported in reference \onlinecite{Jenni2019}.
One expects the spin stiffness to decrease with increasing temperature following the magnetization \cite{Itoh2016,Jenni2019}, but in
SrRuO$_3$ the spin stiffness is identical at the two studied temperatures, which can be seen in the comparable diameter of the ring shaped scattering in Fig. \ref{fig:SRO-TOF-QQ}.
The magnetic signals are however broadened at 160\,K as it is seen in the changed intensity distribution.
The rings of scattering in the constant-energy cuts, see Fig. \ref{fig:SRO-TOF-QQ}(f)-(i), transform towards a more disc-like distribution suggesting
the transformation from long-range magnon into paramagnon scattering.
Within Stoner theory \cite{Lowde1970,Moriya1985} one expects the continuum of magnetic excitations to considerably soften with heating and with the associated
reduction of the magnetization. At low energies, the magnon dispersion in SrRuO$_3$ however changes only little up to 160\,K and
even up to 280\,K when inspecting the single 8\,meV scans shown in reference \cite{Jenni2019}. This strongly supports the
persistence of local magnetization and exchange splitting well above the ferromagnetic phase transition.

\subsection{Polarized experiments with a horizontal magnetic field}

Polarized INS experiments on a ferromagnetic material suffer from the depolarization of the neutron beam that is induced by domains and stray fields. Maintaining
a good neutron polarization is experimentally challenging and requires a large guide field to align domains and to overrule any stray fields of the sample magnetization.
In a previous polarized INS experiment on SrRuO$_3$ on a cold triple-axis spectrometer the feasibility of polarized experiments was demonstrated but these experiments focused on the chirality of the zone-center magnons \cite{Jenni2022}.
In a usual ferromagnet the chirality of this excitation is determined by the right-handedness of the commutation rules
for the components of a spin operator, but it was proposed that the strong spin-orbit coupling in SrRuO$_3$ may result in left-handed excitations \cite{Onoda2008}.
The experiment, however, finds perfect right-handedness \cite{Jenni2022} well in the ferromagnetic phase.

\begin{figure}[htbp]
 \includegraphics[width=0.88\columnwidth]{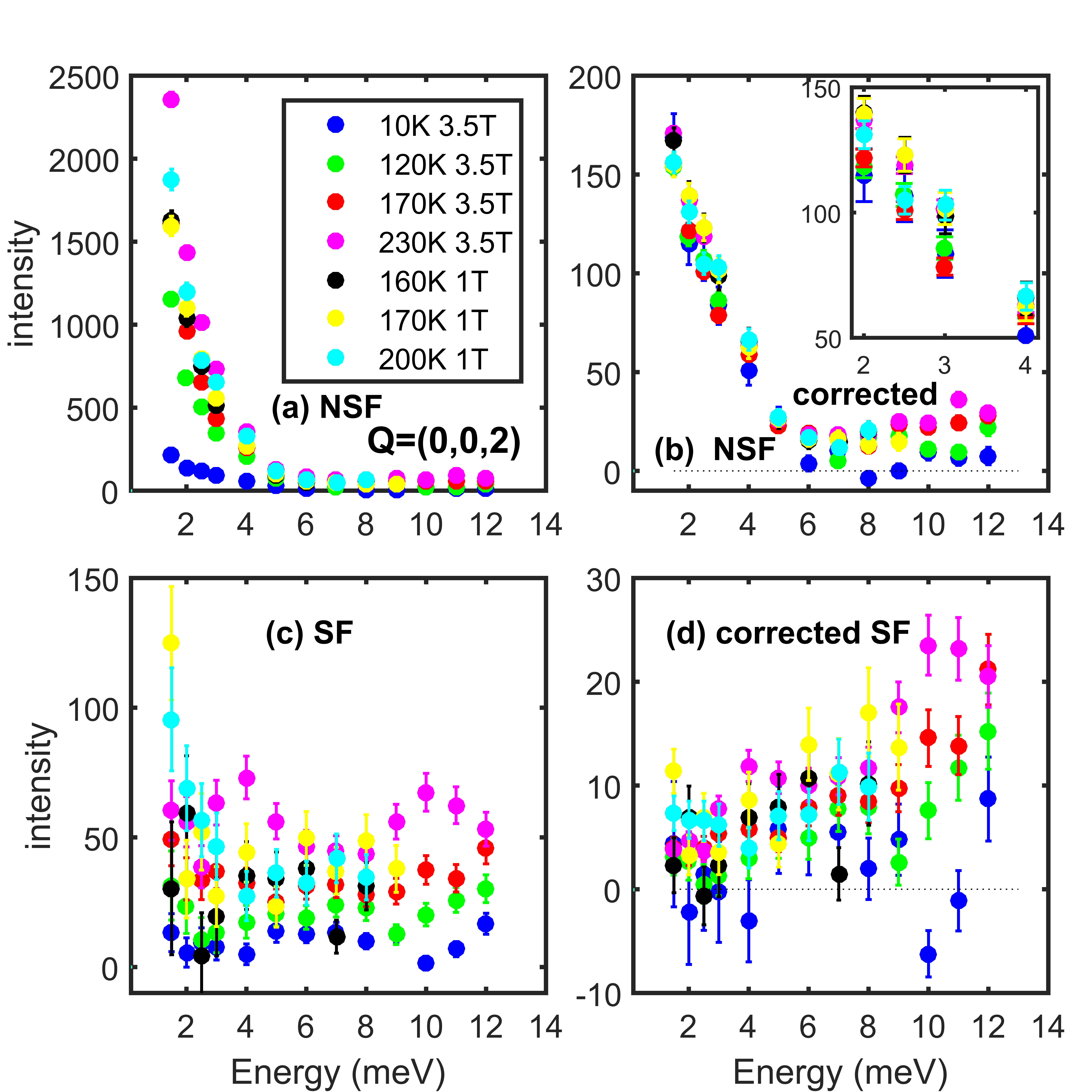}
  \caption{{ Polarized neutron scattering results} for energy scans at the (0,0,2) Bragg point of SrRuO$_3$ performed on the IN20 \tas. Panels (a) and (c) show the NSF and SF intensities corrected for the finite flipping ratios determined at the respective temperature and magnetic field. In panel (b) the NSF data are corrected for the Bose factor and the inset shows a zoom on low energies.
  A small constant background is subtracted from the SF data and a correction for the Bose factor is applied in panel (d). }
  \label{pol-002}
 \end{figure}

\begin{figure}[htbp]
 \includegraphics[width=0.95\columnwidth]{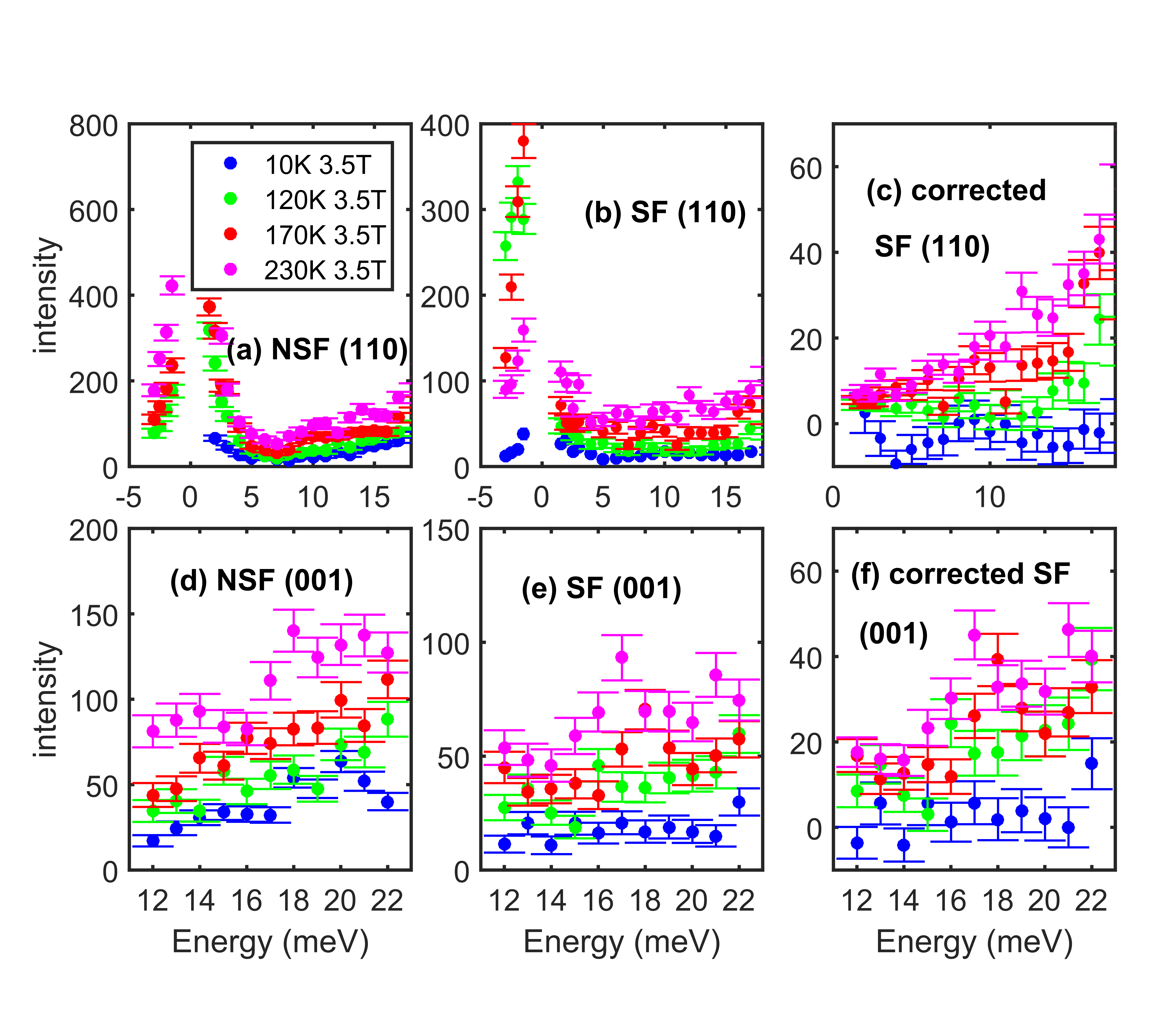}
  \caption{{ Polarized neutron scattering results} for energy scans at the (0,0,1) and (1,1,0) Bragg points of SrRuO$_3$ performed on the IN20 \tas. Panels a(d) and b(e) show the NSF and SF intensities corrected for the finite flipping ratios for the two ${\bf Q}$ values. A small constant background is subtracted from the SF data and a correction for the Bose factor is applied in panels c and f.}
  \label{pol-110-001}
 \end{figure}

With the polarized INS experiment on the thermal spectrometer IN20 we wanted to search for longitudinal modes, i.e. modes with an oscillating moment parallel to the static magnetization in contrast to the transversal character of the magnon modes corresponding to a precession of the moments around the static magnetization. Longitudinal spin excitations were theoretically deduced from
random-phase-approximation calculations \cite{Lowde1970,Moriya1985,Solontsov1995,Solontsov2005} but experimental studies are limited to a few systems and to temperatures close to the magnetic transition, where the  longitudinal response corresponds to critical scattering \cite{Boeni1991,Boeni2002}. 
In our experiment on SrRuO$_3$, the polarization analysis also yields a better separation of magnetic and phonon contributions
at higher temperature, where the magnetic response becomes very broad.
The large sample was mounted in a horizontal magnet cryostat which allows to apply 3.8\,T. In order to
avoid quenching we stayed slightly below this value and applied a magnetic field of 3.5\,T along the [1,1,0] direction, which together with [0,0,1] spans the scattering plane.
Most parts of the experiment were performed by using only the flipper between sample and analyzer whose currents had to be adapted to the stray fields of the horizontal magnet at the flipper position that depend on the angle between the field and the outgoing beam.
The flipper between the monochromator and the sample was only used to verify the polarization at a few points in ${\bf Q}$-E space.
The flipping ratios were measured at the two Bragg reflections (1,1,0) and (0,0,2) to amount to 21.4 and 21.6, respectively, in the paramagnetic state at 230\,K.
However, the quality of the neutron polarization considerably diminishes upon cooling into the ferromagnetic state. At the temperatures of 170, 120  and 10\,K we find the values 19.8[18.8], 11.6[6.8] and 8.3[4.9] at the reflection (1,1,0)[(0,0,2)].
The reduction of the polarization quality is more severe at the (0,0,2) reflection
for which the magnetic field is perpendicular to the scattering vector, so that stray fields of the sample magnetization are more harmful.
For a magnetic field of only 1\,T the flipping ratio is even more rapidly suppressed to 5.8 measured for (0,0,2) at 160\,K. At 10\,K and 3.5\,T
the flipping ratio was also studied on a phonon at (2,2,0.2) yielding a flipping ratio of 8.0 in good agreement with a measurement of the (1,1,0) Bragg
reflection.
Clearly, polarization can be maintained in the ferromagnetic state of SrRuO$_3$ but a careful correction of the reduced
flipping ratios is required and was applied to all data shown in Figures  \ref{pol-002} and  \ref{pol-110-001}.

The horizontal magnet imposes severe restrictions on the accessible angles and it is fixed to the [1,1,0] direction and thereby imposes the
direction of the neutron polarization at the sample. Therefore, it is not possible
to measure the scattering in the usual $x$,$y$,$z$ directions of polarized neutron experiments \cite{Chatterji2005}, but for ${\bf Q}$=(0,0,1) and (0,0,2)
we may only study the spin-flip (SF) $y$ and non-spin-flip (NSF) $y$ channels (i.e. polarization direction perpendicular to the scattering vector
with in the scattering plane) and for ${\bf Q}$=(1,1,0) only SF $x$ and NSF $x$ (i.e. polarization parallel to the scattering vector).
By cooling in a finite field we obtain a nearly monodomain magnetic state with the orthorhombic $c$ direction, the easy axis of magnetic order
in SrRuO$_3$, aligned parallel to the magnetic field.
Since INS only senses magnetic components perpendicular to the scattering vector and since a neutron spin-flip requires a magnetic
component perpendicular to the polarization direction, we obtain the following the selection rules: For ${\bf }Q$=(0,0,1) and (0,0,2),
the NSF signal contains the nuclear scattering and the magnetic excitations polarized parallel to the static magnetization, i.e. longitudinal
excitations, and the SF scattering senses magnetic excitations polarized perpendicular to the magnetization, i.e. transverse magnetic
excitations. For ${\bf Q}$=(1,1,0) the NSF signal contains only the nuclear scattering while the SF signal contains twice the transverse magnetic
excitations. Here we assume that the magnetic excitations in the two transverse channels are identical. Due to the angle constraints it was not
possible to study the low-energy response at (0,0,1) but we had to go to (0,0,2) where the magnetic formfactor already reduces the signal
strength.

Fig. \ref{pol-002} presents the energy scans  obtained at the (0,0,2) Bragg peak. The NSF signal at low energy is fully dominated by the
phonon scattering arising around the strong Bragg reflection.
But the correction for the Bose factor presented in Fig. \ref{pol-002} (b)
reveals an extra weak magnetic scattering near 2 to 3\,meV appearing at the temperatures where the spontaneous part of the magnetization most strongly increases.
Note that the finite fields of 1 and 3.5\,T suppress the sharp transition as the symmetry is already broken by the magnetization induced through
the magnetic field, see reference \onlinecite{Jenni2022}.
This extra scattering visible in Fig. \ref{pol-002} at temperatures near the transition
can be attributed to critical longitudinal scattering near the emergence of the ordered magnetic moment.
This longitudinal signal fully agrees with polarized neutron scattering measurements on ferromagnetic Ni and EuS \cite{Boeni1991,Boeni2002}.
However, deep in the ferromagnetic state there is no evidence for a longitudinal excitation up to 12\,meV and also the uptake of intensity in the NSF
channel at finite temperature most likely stems from phonon and multiphonon processes.
The SF channel at (0,0,2) detects
transverse excitations (polarized along orthorhombic $a$) but again there is no evidence for such scattering at low temperature.
Since the scan path is exactly at the zone center, this is
consistent with the usual picture that the parabolically dispersing magnon is the only signal below the continuum.
However, such transverse magnon scattering appears at higher temperatures and
is consistent with the observations at the other two studied Bragg peaks.

The polarized energy scans at the other two Bragg peaks are shown in Fig. \ref{pol-110-001}. 
At ${\bf Q}$=(1,1,0) the NSF scattering is entirely
due to nuclear scattering and measures the phonon and multiphonon processes. 
There is a very strong phonon at 20\,meV that was also seen in the time-of-flight
data taken on Merlin discussed above. 
The SF scattering at (1,1,0) is flat at low temperature indicating the absence of magnetic scattering in agreement
with the magnon dispersion. 
However, this SF scattering considerably increases with increasing temperature which reflects the above discussed
enhancement of the widths of the magnon signal.
The $Q$ space in the center of the ferromagnetic magnon dispersion gets consecutively filled
with increasing temperature, so that the character of the scattering changes from magnon to paramagnon like.
The consistent observation is also made in the SF scattering at ${\bf Q}$=(0,0,1), see Fig. \ref{pol-110-001} (e,f). The fact that at (1,1,0) we see two transversal magnetic channels is compensated
by the square of the Ru formfactor that is about twice as large at (0,0,1). At low temperature there is a finite signal in the
NSF channel at (0,0,1) that can be safely ascribed to the 20\,meV phonon which has been also observed at (0,0,3) where the
$Q^2$ factor strongly enhances the signal. So there is no evidence for a longitudinal mode up to $\sim$20\,meV.
This agrees with the magnon dispersion extending to at least 35\,meV as one would expect longitudinal excitations
to be strongly suppressed deep in the ferromagnetic phase.
Furthermore, recent ARPES studies indicate that the exchange-induced band energy splitting is rather large in SrRuO$_3$,
of the order of 120\,meV \cite{Hahn2021}, which also implies a larger energy scale for the Stoner continuum and longitudinal modes.

The SF scattering at (1,1,0) has been also measured for negative energy transfer, see Fig. \ref{pol-110-001} (b), where close to the onset of spontaneous magnetization
a strong signal appears at a few meV that has no counterpart at positive energy transfer. This is due to the chirality of the zone-center magnon as it is discussed
in detail in reference \onlinecite{Jenni2022}.
The Heusler polarizing monochromator and analyzer crystals transmit the neutron polarization antiparallel to the guide field. Therefore,
the spin-flip process with the flipper between sample and analyzer turned on and the first flipper being turned off correspond to a scattering from antiparallel to parallel
neutron polarization. A right-handed mode, however, requires the opposite for a positive energy transfer but becomes visible at negative energy transfer as it
is seen in Fig. \ref{pol-110-001} (b). This experiment confirms the perfect right-handedness of the magnon in SrRuO$_3$ \cite{Jenni2022}.

\section{Discussion and Conclusions}

The combined INS study of the magnetic excitations in the ferromagnetic state of SrRuO$_3$ does not reveal the Stoner continuum expected for an itinerant
system. This can be attributed to the still limited energy range for which reliable INS data could be obtained. The magnon modes can be followed up to $\sim$35\,meV but already above the 25\,meV the signal becomes quite reduced. 
In view of the recent ARPES study determining the band-energy splitting to 120\,meV one may expect
the Stoner continuum at comparable energy scales and thus the strongest effects even above the accessible energy range of our experiment. 
Due to the multi-orbital nature of
the electronic band structure the crossover from magnon to Stoner excitations can be more complex in SrRuO$_3$.
The limited ordered moment in SrRuO$_3$ combined with a rapidly decreasing magnetic form factor hamper neutron scattering studies,
so that considerable efforts are needed to cover higher energies. So far, the polarized INS experiments cannot detect any evidence for
longitudinal modes for energies below $\sim$20\,meV in the ferromagnetic state.

The most remarkable feature of the magnetic excitations in SrRuO$_3$ concerns the anomalous temperature dependence of the magnon stiffness and of the gap that
both harden upon heating \cite{Itoh2016,Jenni2019}. These anomalous temperature dependencies follow that of the anomalous Hall effect and can be explained
by the impact of the Weyl points on the spin dynamics. Evidence for Weyl points situated close to the Fermi level has been deduced from DFT calculations \cite{Chen2013}
as well as from magnetotransport studies \cite{Fang2003,Takiguchi2020}. However, the magnon modes remain extremely broad in SrRuO$_3$ even at low temperature.
In order to reproduce the measured data profiles we have to fold the experimental resolution function with a magnon response that shows an energy broadening of 40\% of its energy.
This severe broadening further increases upon heating.
While at low temperature the magnetic response remains essentially magnon like, although exhibiting enormous life-time reduction,
the shape of the magnetic signal changes upon heating.
Close to the magnetic transition the ${\bf Q}$ space in the center of the magnon dispersion surface gets more and more filled which
resembles the intensity distribution of nearly ferromagnetic systems with a paramagnon signal \cite{Moriya1985}.
However the dispersion of the peak energies
is little affected upon heating close to the transition at 160\,K and even well above indicating that the local exchange splitting
remains still considerably larger than the energy range of our experiments.
There are several explanations for the reduced lifetimes of magnons in SrRuO$_3$. The pronounced kink of the electric resistance at
the ferromagnetic transition underlines a strong electron-magnon interaction. In addition the Weyl points and the Berry curvature
imply further scattering paths\cite{Jenni2019} that in view of the strong impact of the topology on the magnon dispersion may also be
important for the magnon damping.

The magnetic excitations have been studied in several metallic ruthenates of the Ruddlesden-Popper series.
In Sr$_2$RuO$_4$ there are dominating incommensurate excitations that arise from pronounced Fermi surface nesting of quasi-one-dimensional sheets \cite{Mazin1999,Sidis1999} and that seem to condense into static incommensurate magnetic order upon minor substitution \cite{Braden2002}.
This nesting is rather robust and can also be observed in Ca$_{2-x}$Sr$_x$RuO$_4$ compounds with $x\sim0.5$ that are closer
to a ferromagnetic instability and that exhibit dominant nearly ferromagnetic magnetic fluctuations\cite{Steffens2011}, see below.
Two recent ARPES studies  yield evidence for flat Fermi-surface sheets \cite{Hahn2021,Lin2021} in SrRuO$_3$ that resemble the
strong nesting in Sr$_2$RuO$_4$. 
The distance of these flat sheets in the [$\xi$ 0 0] direction can roughly be determined to $\xi_{nes}$=0.29 \cite{Hahn2021} and 0.34 \cite{Lin2021} reduced lattice units, respectively, but only reference \onlinecite{Hahn2021} differentiates majority and minority sheets.
The constant energy maps presented in Figures \ref{fig:SRO-TOF-Donuts-lowT} and \ref{fig:SRO-TOF-QQ} yield no indication for such
scattering at either ($\xi_{nes}$ 0 0),  ($\xi_{nes}$ $\xi_{nes}$ 0) or  ($\xi_{nes}$ $\xi_{nes}$ $\xi_{nes}$). 
The strongest nesting peak
in Sr$_2$RuO$_4$ \cite{Mazin1999,Sidis1999} arises along the diagonal profiting from the nesting in two directions, while such an
effect cannot be deduced from the Fermi-surface sheets reported for SrRuO$_3$\cite{Hahn2021,Lin2021}.
In addition to the nesting induced magnetic excitations, Sr$_2$RuO$_4$ also exhibits a broad quasi-ferromagnetic signal \cite{Steffens2019}.
But this  response of Sr$_2$RuO$_4$ is still quite different from the magnon signal that SrRuO$_3$ shows at low temperature.
The response in Sr$_2$RuO$_4$ is little structured in ${\bf Q}$ space and thus approaches a scenario with local interaction that is deduced from DMFT calculations \cite{Strand2019}.

The magnon-like response in SrRuO$_3$ also differs from the quasi-ferromagnetic scattering, which was observed in layered ruthenates that are close to ferromagnetic order.
In Ca$_{2-x}$Sr$_x$RuO$_4$ a ferromagnetic cluster glass ordering is reached for $x$$\sim$0.5 and a metamagnetic transition is formed for further
reduced Sr content \cite{Nakatsuji2000,Nakatsuji2003}. Also Sr$_3$Ru$_2$O$_7$ exhibits a metamagnetic transition and is thus very close to ferromagnetic
order \cite{Perry2001}. The INS studies in Sr$_3$Ru$_2$O$_7$ \cite{Capogna2003} and in Ca$_{2-x}$Sr$_x$RuO$_4$ \cite{Friedt2004,Steffens2011,Steffens2007}
reveal a remarkably similar picture in these layered systems that however differs from the magnon-like response in SrRuO$_3$. The layered materials at
zero field exhibit still incommensurate magnetic fluctuations though appearing at different positions in ${\bf Q}$ space compared to the nesting induced signals
in Sr$_2$RuO$_4$ \cite{Sidis1999,Braden2002a}. The peaks in the magnetic susceptibility of Sr$_3$Ru$_2$O$_7$ and Ca$_{2-x}$Sr$_x$RuO$_4$ appear along
the bond direction and at much smaller absolute values of the propagation vector in agreement with a more ferromagnetic nature. Only for metamagnetic
Ca$_{1.8}$Sr$_{0.2}$RuO$_4$ at finite magnetic field a parabolic and thus magnon-like dispersion was observed \cite{Steffens2007} that finally resembles the magnon dispersion in SrRuO$_3$.
Overall the magnetic response in the layered ruthenates including Sr$_2$RuO$_4$ seems mostly determined by Fermi-surface effects with small
but finite propagation vectors, while SrRuO$_3$ and only the high-field phase of Ca$_{1.8}$Sr$_{0.2}$RuO$_4$ exhibit a parabolic and thus an intrinsic ferromagnetic response.
The response induced by Fermi-surface effects in the layered materials emerges in the form of stacks of scattering in ${\bf Q}-E$ space.
Upon heating the magnetic excitations in SrRuO$_3$, however, approach such a shape.

In conclusion the combined INS study of magnetic excitations in SrRuO$_3$ can characterize a low-temperature magnon dispersion up to rather high
energies that are consistent with a large band energy splitting. Besides the anomalous temperature dependence of the magnon stiffness and of the gap,
the severe broadening of magnons even at low temperature is most remarkable. Upon heating towards the magnetic transition in SrRuO$_3$ this broadening is
further enhanced and finite magnetic response is found at the center of the dispersion. Although the magnon dispersion remains visible up to near
T$_c$ this change indicates an enhanced local character of the interaction and it approaches the findings in other ruthenates where even 
response yields stacks of scattering in $Q$,E space mostly associated with Fermi-surface effects.


\begin{acknowledgments}
This work was funded by the Deutsche Forschungsgemeinschaft (DFG,
German Research Foundation) - Project number 277146847 - CRC 1238, projects A02 and B04.
Experiments at the ISIS Neutron and Muon Source were supported by a beamtime allocation RB1510482 from the Science and Technology Facilities Council.
\end{acknowledgments}


\begin{thebibliography}{73}
\bibitem [{not()}]{note-g}%
	\BibitemOpen
	\href@noop {} {}\bibinfo {note} {The saturation magnetization is proportional
		to the $g$ factor multiplied with the averaged spin. Therefore we can
		determine the averaged spin and use it to extract $J_1$ from $2J_1S$.
		$\langle S_z \rangle$ $M_{sat}=g\langle S_z \rangle \Leftrightarrow
		1.6\,\mu_B =2\langle S_z \rangle \Rightarrow \langle S_z \rangle=0.8$. Using
		the experimental value of $g$ the coupling constant $J_1$ amounts to
		3.4(7)\,meV, because the averaged spin calculated with the saturation
		magnetization increases, $M_{sat}=g\langle S_z \rangle \Rightarrow \langle
		S_z \rangle=1.6/1.78(5)=0.90(3)$.}\BibitemShut {Stop}%
\end{thebibliography}
\end{document}